\providecommand{\adsurl}[1]{\href{#1}{ADS}}

\documentclass[iop,onecolumn]{emulateapj} 
\usepackage{graphics,graphicx,epsf,natbib}
\usepackage{subfigure}
\usepackage[colorlinks=true,linkcolor=blue,anchorcolor=blue,citecolor=blue,urlcolor=blue]{hyperref}
\usepackage{color}
\usepackage{graphicx}
\usepackage{amssymb,amsmath}
\usepackage{natbib}

\slugcomment{Submitted 30 May 2014; Accepted 25 August 2014}

\shorttitle{CLASH-X Mass Profiles}
\shortauthors{Donahue et al.}

\begin{document}


\title{CLASH-X: A Comparison of Lensing and X-ray Techniques for Measuring \\ the Mass Profiles of Galaxy Clusters}


\author{Megan Donahue\altaffilmark{1,2}, 
              G. Mark Voit\altaffilmark{1}, 
              Andisheh Mahdavi\altaffilmark{3},
              Keiichi Umetsu\altaffilmark{4},
              Stefano Ettori\altaffilmark{5},
              Julian Merten,\altaffilmark{6},
               Marc Postman\altaffilmark{7},
               Aaron Hoffer\altaffilmark{1},
              Alessandro Baldi\altaffilmark{1},
              Dan Coe\altaffilmark{7},
              Nicole Czakon\altaffilmark{4},
              Mattias Bartelmann\altaffilmark{15},
              Narciso Benitez\altaffilmark{8},
              Rychard Bouwens\altaffilmark{16},
              Larry Bradley\altaffilmark{7},
             Tom Broadhurst\altaffilmark{9},
              Holland Ford\altaffilmark{10},
              Fabio Gastaldello\altaffilmark{18,23}
              Claudio Grillo\altaffilmark{26}
               Leopoldo Infante\altaffilmark{17},
               Stephanie Jouvel\altaffilmark{10},
               Anton Koekemoer\altaffilmark{7},
               Daniel Kelson\altaffilmark{19},
               Ofer Lahav\altaffilmark{11},
	      Doron Lemze\altaffilmark{7},
               Elinor Medezinski\altaffilmark{7},
               Peter Melchior\altaffilmark{20},
               Massimo Meneghetti\altaffilmark{5,6},
               Alberto Molino\altaffilmark{12},
               John Moustakas\altaffilmark{21},
               Leonidas A. Moustakas\altaffilmark{6}, 
               Mario Nonino\altaffilmark{22},
                Piero Rosati\altaffilmark{13},
               Jack Sayers\altaffilmark{14},
              Stella Seitz\altaffilmark{24},
               Arjen Van der Wel\altaffilmark{25}
               Wei Zheng\altaffilmark{10},
               Adi  Zitrin\altaffilmark{14,27},
              }

\altaffiltext{1}{Physics and Astronomy Dept., Michigan State University, East Lansing, MI, 48824 USA}
\altaffiltext{2}{donahue@pa.msu.edu }
\altaffiltext{3}{San Francisco State University, San Francisco, CA USA}
\altaffiltext{4}{Institute of Astronomy and Astrophysics, Academia Sinica, Roosevelt Rd, Taipei 10617, Taiwan}
\altaffiltext{5}{Osservatori Astronomico di Bologna, Via Ranzani 1, 40127 Bologna, Italy; INFN, Sezione di Bologna, viale Berti Pichat 6/2, I-40127 Bologna, Italy}
\altaffiltext{6}{Jet Propulsion Laboratory, 4800 Oak Grove Dr., Pasadena, CA 91109 USA}
\altaffiltext{7}{STScI,3700 San Martin Drive, Baltimore, MD 21218 USA}
\altaffiltext{8}{Instituto de Astrofisica de Andalucia (CSIC), C/Camino Bajo de Hu{\'e}tor 24, Granada 18008, Spain}
\altaffiltext{9}{Dept of Theoretical Physics, University of the Basque Country, 48080 Bilbao, Spain}
\altaffiltext{10}{Dept of Physics \& Astronomy, The Johns Hopkins University, Baltimore, MD, 21218 USA}
\altaffiltext{11}{Dept of Physics \& Astronomy, University College London, Gower St., London, WC1E 6BT}
\altaffiltext{12}{Instituto de Astrofisica de Andalucia, Glorieta de la astronomia s/n, 18008. Granada Spain}
\altaffiltext{13}{Dept of Physics and Astronomy, University of Ferrara, Via Saragat, 1-44122 Ferrara, Italy} 
\altaffiltext{14}{Dept of Astronomy, California Institute of Technology, 1200 E. California Blvd., Pasadena, CA 91125 USA}
\altaffiltext{15}{Universit{\"a}t Heidelberg, Zentrum f{\"u}r Astronomie, Philosophenweg 12, 69120 Heidelberg, Germany}
\altaffiltext{16}{Leiden Observatories, Niels Bohrweb 2, NL-2333 CA Leiden, The Netherlands}
\altaffiltext{17}{Dept Astronom{\'i}a-Astrof{\'i}sica, Pontificia Universidad Cat{\'o}lica de Chile, Casilla 306, 22 Santiago, Chile}
\altaffiltext{18}{ INAF-IASF, via Bassini 15, 20133 Milan, Italy}
\altaffiltext{19}{Carnegie Observatories, 813 Santa Barbara Street, Pasadena, CA 91101 USA}
\altaffiltext{20}{Center for Cosmology and Astroparticle Physics, The Ohio State University, Columbus, OH USA}
\altaffiltext{21}{Dept of Physics and Astronomy, Siena College, Loudonville, NY USA}
\altaffiltext{22}{INAF-Osservatori Astronomico di Trieste, 34143, Trieste, Italy}
\altaffiltext{23}{University of California at Irvine, 4129 Frederick Reines Hall, Irvine, CA 92697-4575, USA}
\altaffiltext{24}{Dept. f{\"u}r Physik, Universit{\"a}ts-Sternwarte M{\"u}nchen, Scheinerstrasse 1, D-81679 M{\"u}nchen, Germany}
\altaffiltext{25}{Max Planck Institue for Astronomy, K{\"o}nigstuhl 17, D-69117, Heidelberg, Germany}
\altaffiltext{26}{Dark Cosmology Centre, Niels Bohr Institute, University of Copenhagen, Juliane Maries Vej 30, DK-2100 Copenhagen, Denmark}
\altaffiltext{27}{Hubble Fellow}

\begin{abstract}

\vspace{1em}
We present profiles of temperature, gas mass, and hydrostatic mass estimated from new and archival X-ray observations of CLASH clusters. We compare measurements derived from XMM 
and {\em Chandra} observations with one another and compare both to gravitational lensing mass profiles derived with CLASH Hubble Space Telescope and Subaru Telescope lensing data. Radial profiles of   
{\em Chandra} and XMM measurements of electron density and enclosed gas mass are nearly identical, indicating that differences in hydrostatic 
masses inferred from X-ray observations arise from differences in gas-temperature measurements. Encouragingly, 
gas temperatures measured in clusters by XMM and {\em Chandra} are consistent with one another at $\sim 100-200$~kpc radii 
but XMM temperatures systematically decline relative to 
{\em Chandra} temperatures at larger radii. The angular dependence of the discrepancy suggests additional investigation on systematics such
as the XMM point spread function correction, vignetting and off-axis responses is yet required.
We present the CLASH-X mass-profile comparisons in the form of cosmology-independent and redshift-independent circular-velocity profiles.   
We argue that comparisons of circular-velocity profiles are the most robust way to assess mass bias.
Ratios of {\em Chandra} HSE mass profiles to CLASH lensing profiles show no obvious radial dependence in the 0.3--0.8 Mpc range.  However, the mean mass biases inferred from the WL and SaWLens data are different. As an example, the weighted-mean value at 0.5 Mpc is $\langle b \rangle = 0.12$ for the WL comparison and $\langle b \rangle = -0.11$ for the SaWLens comparison.
The ratios of XMM HSE mass profiles to CLASH lensing profiles show a pronounced radial dependence in the 0.3--1.0 Mpc range, with a weighted mean mass bias of value rising to 
$\langle b \rangle \gtrsim 0.3$ at $\sim 1$~Mpc for the WL comparison and $\langle b \rangle \approx 0.25$ for the SaWLens comparison. The enclosed gas mass profiles from both {\em Chandra} and XMM rise to a value $\approx 1/8$ times the total-mass profiles inferred from lensing at $\approx 0.5$~Mpc and remain constant outside of that radius, suggesting that 
$M_{\rm gas} \times 8$ profiles may be an excellent proxy for total-mass profiles at $\gtrsim 0.5$~Mpc in massive galaxy clusters.
\vspace{1em}
 
\end{abstract}


\section{Introduction}
\setcounter{footnote}{0}

CLASH \citep{2012ApJS..199...25P} is a {\em Hubble} Multi-Cycle Treasury program to observe massive galaxy clusters at intermediate redshifts.  It has three major scientific goals:  (1) to compare the observed properties of galaxy clusters with the predictions of $\Lambda$CDM cosmology, (2) to search for galaxies at redshift $z \sim 10$ using massive clusters as gravitational lenses, and (3) to discover and monitor distant supernovae in the cluster fields through a staggered program of multi-wavelength observations.  
HST observations of CLASH clusters have enabled the discoveries of $z > 10$ lensed galaxies \citep{2013ApJ...762...32C,2012Natur.489..406Z} and numerous multiply-lensed  galaxies at $z=4-7$ \citep{2012ApJ...747L...9Z,2013A&A...559L...9B,2014MNRAS.438.1417M}.
CLASH has also produced strong-lensing analyses for the inner portions of individual galaxy clusters \citep{2012ApJ...749...97Z} 
and combined weak-lensing/strong-lensing analyses \citep{2012ApJ...755...56U,2012ApJ...757...22C,2013ApJ...774..124E,2013ApJ...777...43M}. 
The supernova survey component of CLASH discovered 39 supernova candidates and used 27 of them, discovered in the parallel fields, to measure the type Ia supernova rate out to $z \sim 2$ \citep{2014ApJ...783...28G}. Moreover, three of the supernovae discovered in the prime fields were lensed by the galaxy clusters, as reported  in \citet{2014ApJ...786....9P}.
Umetsu et al. (2014) describes the weak lensing analyses of a subsample the CLASH clusters, while Merten et al. (2014) has produced a simultaneous analysis of the strong and weak lensing of a slightly different CLASH subsample.

This paper focuses on the properties of the CLASH clusters themselves, and particularly on the cluster masses and mass profiles derived from X-ray and gravitational-lensing observations \citep{2014arXiv1404.1375U,2014arXiv1404.1376M}.  According to the $\Lambda$CDM model, the gravitational influence of the invisible particles we call dark matter should produce intricate and beautiful networks of large-scale filaments with massive clusters of galaxies at the intersections. In the deep potential wells of galaxy clusters, diffuse intergalactic gas reaches temperatures sufficient to radiate X-ray light, unveiling  rare and distant massive structures  \citep[e.g.][]{2004JCAP...09..011P,1990ApJS...72..567G}.
  The implications of the predominance of the hot gas in the cluster baryon budget have been known for a while \citep[e.g.][]{1971ApJ...169L..13G,1991ApJ...372..410H,1995ApJ...445..578D}, but progress was slow until 
large X-ray surveys, especially the ROSAT All-Sky Survey, revealed hundreds of massive clusters out to redshifts $>0.5$ 
\citep[e.g.][]{2001ApJ...553..668E,1998ApJ...492L..21R}. Many workers since have observed these clusters with {\em Chandra} and {\em XMM-Newton} to derive important cosmological constraints on dark matter, dark energy, and the baryonic mass fractions of clusters from analyses of this cluster population \citep{2009ApJ...692.1033V,2002MNRAS.334L..11A,2010MNRAS.406.1759M,2009A&A...501...61E,2003ApJ...590...15V}.  All such studies require accurate calibrations of the X-ray observables \citep[e.g.][]{2010ApJ...721..875O, 2013MNRAS.435.1265E, 2007A&A...474L..37A,2007A&A...467..437Z} used to measure galaxy-cluster masses \citep{1996ApJ...469..494E}.

CLASH has been testing the $\Lambda$CDM model by measuring the radial mass profiles of clusters with gravitational-lensing observations and comparing them with the simulated profiles of $\Lambda$CDM clusters.  
We have found excellent agreement between the observed mass profiles (Merten et al. 2014, Umetsu et al. 2014) 
and those predicted from a sample of simulated clusters selected in a similar manner from a $\Lambda$CDM  
simulation \citep{2014arXiv1404.1384M}. 
Galaxy clusters in the CLASH sample are generally well fit by an NFW profile \citep{1997ApJ...490..493N} with a halo concentration $c_{200} \equiv r_{\rm s}/r_{200}$, defined in terms of $r_{\rm s}$, the ``scale radius" at which the local slope of the matter-density profile is $\rho \propto r^{-2}$, and $r_{200}$, the radius within which the mean density is 200 times the critical density of the universe, $\rho_{\rm cr}(z)$, at the cluster's redshift.  Concentration values for CLASH clusters are typically $c_{200} \approx 4$.  Both the relationship between a cluster's mass and density-profile concentration found by CLASH and the evolution of that relationship are consistent with $\Lambda$CDM simulations (Merten et al. 2014; Meneghetti et al. 2014).

Here we take advantage of the unparalleled gravitational-lensing data collected by the CLASH collaboration to assess the level of agreement between the cluster mass profiles inferred from X-ray observations and those measured through lensing.
Gravitational lensing is considered the gold standard for cluster mass measurements because of its lack of sensitivity to cluster 
astrophysics \citep{1990ApJ...349L...1T,1993ApJ...404..441K}.  
However, lensing measurements suffer from intrinsic scatter owing to statistical fluctuations in the amount of matter along the line of sight to the cluster but outside of the cluster itself \citep{2005ApJ...622...99D,2007MNRAS.374L..37K}.  
In this paper, as in most cosmological studies of galaxy clusters, we will define the outer boundary of a cluster as the radius $r_\Delta$ of a sphere encompassing a mean matter density $\Delta \rho_{\rm cr}$.  
Projected matter fluctuations outside of this radius do not significantly bias the cluster mass measurements from gravitational lensing but are expected to lead to significant ($\sim 20$\%) scatter between lensing-inferred masses and spherical-overdensity masses
\citep{2001ApJ...547..560M,2001A&A...370..743H,2003MNRAS.339.1155H,2011ApJ...740...25B}.

Cluster mass measurements inferred from X-ray observations are believed to have less statistical scatter than those inferred from lensing but are subject to greater astrophysical uncertainties, which have the potential to introduce systematic bias.   This paper focuses on mass measurements invoking the assumption that the intracluster medium is in hydrostatic equilibrium (HSE), but not all X-ray studies adopt that assumption.  Some rely on the relation between total mass and the best-fitting spectroscopically-determined gas temperature $T_X$, which can be calibrated with either numerical simulations or lensing observations \citep{1999ApJ...520...78H,2001A&A...368..749F,2001ApJ...553...78X,2002A&A...391..841E,2006ApJ...640..691V}.  Others use total gas mass $M_{\rm gas}$ or the quantity $Y_X = T_X \times M_{\rm gas}$ as mass proxies \citep{2007ApJ...668....1N,2010A&A...517A..92A,2010ApJ...721..875O,2013ApJ...767..116M}.   

So why bother with hydrostatic methods, given that galaxy clusters are unlikely to be in perfect hydrostatic equilibrium? 
One answer is that HSE mass estimates are relatively easy to determine for galaxy clusters with a single central peak in X-ray surface brightness, as long as the data are sufficient to generate radial profiles of gas temperature $T_{\rm gas}$ and electron density $n_e$.  
X-ray observatories accomplish this task by collecting X-ray photons while recording each event's energy and point of origin in the sky. Photon events then can be compiled into both 2-d maps of X-ray emission and 1-d spectra of specific regions on the sky, from which $T_X(r)$ and $n_e(r)$ can be determined.
Another answer is that many of the clusters produced in numerical simulations are not far from equilibrium. 
By the late 1990s, cluster simulations of gas outside of the cooling core ($r \gtrsim 100$ kpc) could generate fairly reliable representations of cluster X-ray observations. Those same simulations predicted that non-thermal motions, such as turbulence and bulk motions remaining in the gas after interactions and mergers, provide on average only 10-20\% of the pressure support in the intracluster gas \citep{1990ApJ...363..349E}.
The timescale for intracluster gas to respond to changes in the cluster potential is $\sim T_8^{-1/2} r_{\rm Mpc} \, \rm{Gyr}$,
where $T_8 = T_{\rm gas} / 10^8 K$ and $r_{\rm Mpc}$ is the cluster radius in Megaparsecs. 
Incomplete thermalization of residual motions from a merger event may contribute significant pressure support shortly after a merger, but the cluster subsequently relaxes toward hydrostatic equilibrium on a $\sim 1 \, {\rm Gyr}$ timescale.

While a cosmologist might be concerned about time-varying offsets between a cluster's HSE mass and its actual mass, a true astrophysicist finds them interesting, because the {\em difference} between those masses reflects the thermalization state of gas motions in the intracluster medium and the speed at which they damp.  (See  \citet{2013SSRv..177..195R} for a review of cluster outskirts.)
For example, the presence of long-lived turbulence in a cluster would indicate that the gas has relatively low viscosity, a hydrodynamical property of the cluster gas that is currently not well constrained by observations \citep{2005MNRAS.364..753D,2006MNRAS.371.1025S}.
Measuring systematic differences between the HSE mass of a cluster and the ``true" mass measured by gravitational lensing, also known as the X-ray mass bias $b_X = 1 - M_{\rm HSE} / M_{\rm true}$ \citep{2006ApJ...640..691V,2008ApJ...672..122E}, therefore provides valuable information about the physics of the intracluster medium. 
Theoretical simulations generally predict $\langle b_X \rangle \approx 0.2$ for an unbiased population of clusters \citep[e.g.,][]{2012NJPh...14e5018R,2014ApJ...782..107N}. On a cluster-by-cluster basis, the assumption of spherical symmetry produces scatter in the $b_X$ measurement, because departures from symmetry (or triaxiality) affect the lensing mass estimate more than the X-ray mass estimate \citep[cf.][]{2012ApJ...757...22C}.  Lensing masses have measurement uncertainties of their own, and current estimates of the systematic uncertainties range from about 15-20\% for discrepant measurements for the same cluster from different groups to 8\% for internal systematics for the CLASH weak lensing estimates (Umetsu et al. 2014). 

Lately, interest in the value of $\langle b_X \rangle$ has heightened because of the discrepancy between the cosmological parameters inferred from {\em Planck} observations of the primary anisotropies of the cosmic microwave background (CMB) and the galaxy-cluster counts provided by {\em Planck} observations of the Sunyaev-Zeldovich (S-Z) effect \citep{2013A&A...550A.129P}.   According to the $\Lambda$CDM model, the number density of galaxy clusters of a given mass at a given redshift depends sensitively on the matter-density parameter $\Omega_{\rm M}$ and the matter-perturbation amplitude $\sigma_8$.  Relating $\Lambda$CDM cluster predictions to S-Z cluster counts currently requires an assumption about mass bias because there is no definitive calibration of the relationship between S-Z signal and cluster mass.  Instead, masses determined from X-ray observations with the XMM satellite were used to establish this relationship \citep{2010A&A...517A..92A}, but using the $\langle b_X \rangle \approx 0.2$ value found in simulations in that calibration leads to an overprediction of S-Z counts.  A larger mass bias corresponding to $\langle b_X \rangle \approx 0.4$ can reconcile the S-Z counts with the {\em Planck} CMB cosmology \citep{2013A&A...550A.131P}.   Alternatively, streaming of neutrinos with a mass sum $\sim 0.5 \, {\rm eV}$ can suppress the cosmic perturbations on galaxy-cluster scales enough to explain the discrepancy \citep{2013JCAP...10..044H,2014PhRvL.112e1302W,2014PhRvL.112e1303B}, but this mass value is in tension with other neutrino-mass constraints from large-scale structure.  Either way, accurate measurements of mass bias will be critical for resolving this issue.

In another joint analysis of CMB results and cluster properties, \citet{2014MNRAS.438...78R} 
derive an $L_X-M$ relation that reconciles the thermal SZ power spectrum from 
WMAP7 with cluster scaling relations between richness and mass and the S-Z signal. 
They show rough consistency of this relation with published values $M_{500}$ for 2 CLASH clusters and 
make predictions based on $L_X$ for the masses of a subset of the CLASH clusters.
 
Our discussion of these topics proceeds like this:  Section~2 introduces the CLASH cluster sample.   Section~3 describes the X-ray data analysis.  Section~4 presents the radial profiles of gas density and gas temperature, on which the rest of the analysis is based, and calls attention to a systematic radially-dependent discrepancy between the temperatures measured with XMM and those measured with {\em Chandra}.  Section~5 shows the individual cluster mass profiles as plots of circular velocity $v_{\rm circ} \equiv [GM(r)/r]^{1/2}$ as a function of radius and encourages others to provide mass profiles in this form, in order to minimize dependences on cosmological assumptions.  Section~6 compares the {\em Chandra} mass profiles with the CLASH lensing profiles;  Section~7 does the same for the XMM profiles, showing that the XMM mass bias at large radii may be as great as $\langle b_X \rangle \approx 0.4$.  Section~8 shows that gas masses derived from both {\em Chandra} and XMM closely agree and are quite close to 1/8 of the lensing mass outside of $\sim 0.5$~Mpc, implying that $8 M_{\rm gas}$ is a good approximation to the total enclosed mass within large radii for massive clusters .  Section~9 discusses the implications of our results for the {\em Planck} cluster-mass discrepancy, and \S~10 summarizes our findings.  When necessary, we adopt a vanilla $\Lambda$CDM cosmology with a single decimal place ($\Omega_{\rm M}=0.3$, $\Omega_\Lambda=0.7$, and $H_0=70 \, h_{70}$ km s$^{-1}$ Mpc$^{-1}$) but prefer to state results in forms that depend as little as possible on cosmological assumptions.

\section{The CLASH Cluster Sample}

The CLASH program and strategy are completely described in \citet{2012ApJS..199...25P}.   All 25 clusters are high-mass, high-temperature ($T_X \gtrsim 6$ keV) clusters of galaxies. These 25 galaxy-cluster targets fall into two general categories.  Twenty were chosen for their relatively symmetric X-ray appearance, primarily to test the $\Lambda$CDM predictions for radial mass profiles.  Five are ``high-magnification" clusters chosen for the size of their Einstein radius, primarily to maximize the probability of finding highly-magnified $z > 10$ background galaxies.  Table~\ref{table:CLASH_cluster_data} lists all 25 CLASH clusters, with the high-magnification subset at the bottom, and indicates the data included in this paper for each cluster.  The CLASH weak-lensing result (WL) is based on a joint shear and magnification analysis of datasets primarily from {\em Subaru} and is described in Umetsu et al. (2014).  The CLASH joint modeling of strong- and weak-lensing ---SaWLens--- data from both {\em Hubble} and ground-based data for a subsample of CLASH clusters is described in Merten et al. (2014). Both lensing mass measurements are based on spherical NFW fits to the projected mass density profiles $\Sigma( R)$ recovered from the respective data sets. These features distinguish these analyses from the majority of previous lensing mass measurements, which are mostly based on tangential shear fitting. 

Archival {\em Chandra} X-ray data were available for all the CLASH clusters at the start of the project, because those observations were used to select the targets.  In this regard, the CLASH collaboration is indebted to many other X-ray observers, and especially the Massive Cluster Survey (MACS) led by Harald Ebeling \citep{2001ApJ...553..668E}.  That survey originally identified many of the clusters studied here and collected much of the {\em Chandra} X-ray data.  All of the archival data were sufficient for deriving electron density and gas temperature in a minimum of three radial bins, and we have previously provided independent analyses of many of those datasets in the ACCEPT database \citet{2009ApJS..182...12C}.  

A majority of the CLASH clusters also had archival XMM data, but more than a few of those datasets turned out to be highly contaminated by flares and are unusable for this project.   We therefore acquired new XMM data for six CLASH clusters Abell 2261, MACS1931, MACS1115, MACS0429, MACS1720, and MACS1423 and present those data in this paper. The latter two datasets, especially for MACS1423, were severely compromised by flares, but still represent independent X-ray measurements.  
XMM observations are particularly desirable because of its larger collecting area and field of view, which allow us to extend the X-ray hydrostatic mass profiles out to $r_{500}$, increasing the radial range over which the X-ray and weak-lensing data sets overlap.  

Table~\ref{table:xrayobs} lists both the {\em Chandra} and XMM datasets used in this paper, along with the flare-free exposure times available in each dataset.  We did not utilize all of the observations available in the archive for each cluster because the flare 
contamination could be considerable for some of these datasets, and for others, the gain achieved by adding an incremental amount of exposure was not worth the added systematic uncertainty of potential calibration variance. In general, if a single dataset added fewer than 30\% to the total counts we did not use it.

\begin{deluxetable}{lrrrcrcccc}
\tabletypesize{\scriptsize}
\tablecaption{CLASH Cluster Data \label{table:CLASH_cluster_data}}
\tablewidth{0pt}
\tablehead{
\colhead{Name} & \colhead{RA (J2000)} & \colhead{Dec (J2000)} & \colhead{z} & \colhead{Hi-Mag\tablenotemark{a}} & 
  \colhead{$K_0$\tablenotemark{b}} &
  \colhead{{\em Chandra}} & \colhead{XMM} & 
  \colhead{{\em Subaru} WL\tablenotemark{c}} & \colhead{SaWLens\tablenotemark{d}}  
}
\startdata
 & & & & & & & & & \vspace*{-0.5em}  \\
Abell  209 	   &   01:31:52.54    	&  -13:36:40.4  &   	0.206   &       &   105.5  &  Y   &  Y  &  Y  &  Y    \\
Abell  383		   &   02:48:03.40    	&  -03:31:44.9  &  	0.187   &       &     13.0  &  Y   &  Y  &  Y  &  Y    \\
MACSJ0329-02  &   03:29:41.56    	&  -02:11:46.1  &  	0.450   &       &     11.1  &  Y   &       &  Y  &  Y    \\
MACSJ0429-02  &   04:29:36.05    	&  -02:53:06.1  &	0.399   &       &     17.2   &  Y   &  Y  &  Y  &  Y    \\
MACSJ0744+39 &	07:44:52.82   	& +39:27:26.9  &	0.686   &       &     42.4  &  Y   &  Y   &  Y  &  Y    \\
 & & & & & & & & & \vspace*{-0.5em}  \\ 
Abell  611		   &   08:00:56.82    	& +36:03:23.6  &	0.288   &       & 125~~   &  Y   &        &  Y  &  Y    \\
MACSJ1115+01 &	11:15:51.90	& +01:29:55.1  &	0.355   &       &     14.8  &  Y   &  Y   &  Y  &  Y    \\
Abell1423   	   &   11:57:17.36	& +33:36:37.5 &	0.213   &       &    68.3   &  Y   &        &      &        \\
MACSJ1206-08  &	12:06:12.09	& -08:48:04.4  & 	0.439   &       &    69.0   &  Y   &  Y   &  Y  &  Y    \\
CLJ1226+3332   &	12:26:58.25	& +33:32:48.6 &	0.890   &       &  166~~  &  Y   &  Y/N\tablenotemark{e}   &      &  Y    \\
 & & & & & & & & & \vspace*{-0.5em}  \\
MACSJ1311-03  &	13:11:01.80	& -03:10:39.8  & 	0.494   &       &    47.4    &  Y   &        &      &  Y    \\
RXJ1347-1145   &	13:47:30.62	& -11:45:09.4  & 	0.451   &       &    12.5   &  Y   &    Y   &  Y  &  Y    \\
MACSJ1423+24 &	14:23:47.88	& +24:04:42.5 &	0.545   &       &    10.2   &  Y   &         &       &  Y    \\
MACSJ1532+30 &	15:32:53.78	& +30:20:59.4 &	0.362   &       &    16.9   &  Y   &    Y   &  Y  &  Y    \\
MACSJ1720+35 &	17:20:16.78	& +35:36:26.5 &	0.387   &       &    94.4   &  Y   &         &   Y  &  Y    \\
 & & & & & & & & & \vspace*{-0.5em}  \\  
Abell 2261	   &    17:22:27.18 	& +32:07:57.3 &	0.224   &       &    61.1   &  Y   &    Y   &  Y  &  Y    \\
MACSJ1931-26  &	19:31:49.62	& -26:34:32.9  &	0.352   &       &    14.6   &  Y   &   Y   &   Y   &  Y    \\
RXJ2129+0005  &	21:29:39.96	& +00:05:21.2 &	0.234   &       &    21.1   &  Y   &   Y   &   Y   &  Y    \\
MS2137-2353	   &    21:40:15.17 	& -23:39:40.2  &	0.313   &       &    14.7   &  Y   &   Y   &   Y   &  Y    \\
RXCJ2248-4431   &	22:48:43.96	& -44:31:51.3  & 	0.348   &       &    42.2   &  Y   &   Y   &   Y   &  Y    \\
 & & & & & & & & & \vspace*{-0.5em}  \\  
 \hline
 & & & & & & & & & \vspace*{-0.5em}  \\  
MACSJ0416-24  &   04:16:08.38    	&  -24:04:20.8  &	0.397\tablenotemark{f}   &  Y  &  400~~  &  Y   &       &  Y  &        \\
MACSJ0647+70 &   06:47:50.27    	& +70:14:55.0  &	0.584   &  Y  & 225~~   &  Y   &  Y   &  Y  &       \\
MACSJ0717+37 &   07:17:32.63    	& +37:44:59.7  &	0.548   &  Y  &  220~~  &  Y   &  Y   &  Y  &       \\
MACSJ1149+22 &	11:49:35.69	& +22:23:54.6 &	0.544   &  Y  &  280~~  &  Y   &        &  Y  &        \\
MACSJ2129-07  &	21:29:26.06	& -07:41:28.8  & 	0.570   &  Y  &  200~~  &  Y   &         &        &      \vspace*{-0.5em} \\
\enddata
\tablenotetext{a}{High-Mag of Y indicates the cluster is one of the 5 CLASH clusters selected for their lensing properties.}
\tablenotetext{b}{Core entropy in keV cm$^2$, from the ACCEPT database \citep{2009ApJS..182...12C} unless otherwise noted.  Values of $K_0 \lesssim 30 \, {\rm keV \, cm}^2$ indicate the presence of a strong cool core.}
\tablenotetext{c}{Umetsu et al. 2014. Note: the weak lensing data RXJ2248-4431 is from the 2.2m Wide Field Imager ESO/Chile \citep{2013MNRAS.432.1455G}.}
\tablenotetext{d}{Merten et al. 2014.}
\tablenotetext{e}{XMM data exist for this cluster and were analyzed for CLASH-X but are not of sufficient quality for comparisons with the other data sets.}
\tablenotetext{f}{Redshift for MACS0416-24 from Postman et al. (2012) is updated by \citet{2014ApJS..211...21E}. }
\end{deluxetable}

\begin{deluxetable}{lccccrrrrrr}
\tabletypesize{\scriptsize}
\tablecaption{X-ray Observations\label{table:xrayobs}}
\tablewidth{0pt}
\tablehead{
\colhead{Name} & \colhead{$N_H$\tablenotemark{a}} &
\colhead{XMM} & \colhead{$t_{MOS1}$} & \colhead{$t_{MOS2}$} & \colhead{$t_{pn}$} & 
\colhead{{\em Chandra}} & \colhead{$t_{Ch}$}  \\
\colhead{} &  \colhead{($10^{20} \, {\rm cm}^2$)} &
\colhead{ObsID} & \colhead{(s)} & \colhead{(s)} &\colhead{(s)} &
 \colhead{ObsID} & \colhead{(s)}  }
\startdata
& & & & & & & \vspace{0.5em} \\	
Abell  209	 		&	1.68 &	0084230301	&	16832 	&	16838 	&	11219 	&       522 		& 9962 \\
		 		&		&			    	&		   	&		   	&			& 	3579 	& 9935 \\
& & & & & & & \vspace{-0.5em} \\	
Abell  383	 		&  	4.07 &	0084230501	&	23089 	&	25071 	&	20237 	&	2321		& 19509 \\
		 		&		&			 	&		   	&		   	&			&	2320 	& 19259 \\
& & & & & & & \vspace{-0.5em} \\	
MACSJ0329-02	&	6.21	&				&	    		&       		&       		&   	3582 	& 19850 \\
				&		&				&			&			&			&	6108 	& 39644 \\
& & & & & & & \vspace{-0.5em} \\	
MACSJ0429-02	&	5.7	&	0720700101	&     93998	&  97125	&  77609	&  	3271 	& 23166  \\
& & & & & & & \vspace{-0.5em} \\	
MACSJ0744+39	&	4.66	&	0551850101	&	29556	&	33507	&	14636	&  	3197 	& 20238 \\
                			&          	&   	0551851201	& 	44197	&  	47184 	&  	35476	&   	3585 	& 19604\\
                 		&          	&        			&			&			&			&	6111 	& 40430 \\
& & & & & & & \vspace{0.5em} \\	
Abell  611		&  	4.99	&	0605000601\tablenotemark{b}	&	6855 	&	6933 	&	1598 	&   	3194 	& 36113 \\
& & & & & & & \vspace{-0.5em} \\	
MACSJ1115+01	&	4.14	&	0693180201	&	17807	&	17809	&	11531 	&  	9375 	& 39326\\
& & & & & & & \vspace{-0.5em} \\	
Abell1423   		&   	1.00	&           			&       		&        		&      			&   11724 	& 25705 \\
& & & & & & & \vspace{-0.5em} \\	
MACS1206-08	&	4.15	&  0502430401	& 29183	&  29203	& 21040	&   	3277 	& 23258 \\
& & & & & & & \vspace{-0.5em} \\	
CLJ1226+3332	&	1.37	&	0070340501\tablenotemark{b}	&	11604	&	10615	&	6852  	&  	3180 	& 31301\\
				&		&	0200340101\tablenotemark{b}	&  	47422	&   	46467	&   	32806	&	5014 	& 32444\\
& & & & & & & \vspace{0.5em} \\	
MACSJ1311-03	&	2.18	&				&	     		&      			&      			&       6110 	& 63206 \\
& & & & & & & \vspace{-0.5em} \\	
RXJ1347-1145		&	4.89	&	112960101	&	28743	&	29727	&	21712	&   	3592 	& 57319\\
				&		&				&			&			&			&	13999 	& 53611\\
				&		&				&			&			&			&	14407 	& 62847 \\
& & & & & & & \vspace{-0.5em} \\	
MACSJ1423+24	&	2.65	&	0720700301\tablenotemark{b}     &	28635   & 31217   & 14300     	&    	4195 	& 115551\\
               &		    &	0720700401						&	23428   & 22419   & 19696      		&    	   	 	& 		\\
& & & & & & & \vspace{-0.5em} \\	
MACSJ1532+30	&	2.21	&	0039340101	&	8906 	&	7917	&	7184		&   14009 	& 88662\\
& & & & & & & \vspace{-0.5em} \\	
MACSJ1720+35	&	3.35	&	0720700201	&	22775   & 31690 	& 16100 	&   6107 	& 33489\\
				&		    &				&			&			&			&	3280 	& 20811\\
& & & & & & & \vspace{0.5em} \\	
Abell 2261		&    	3.31&   0693180901	&	28323	&	29339	&	23223	&   	5007 	& 24317 \\
& & & & & & & \vspace{-0.5em} \\	
MACSJ1931-26	&	9.31	&	0693180101	&	39222	&	39242	&	33903 	&  	9382 	& 98922\\
& & & & & & & \vspace{-0.5em} \\
& & & & & & & \vspace{-0.5em} \\	
RXJ2129+0005	&	4.3	&   	0093030201	&	30359	&	32358	&	23284	&   	9370 	& 29635  \\
	
& & & & & & & \vspace{-0.5em} \\	
MS2137-2353		&    	3.4	&   	0673830201	&	55329	&	53362	&	37106	&   	4974 	& 24440\\
				&		    &				&			&			&			    &	   5250 	& 34069\\
& & & & & & & \vspace{-0.5em} \\	
RXCJ2248-4431		&	1.77	&	0504630101	&	24695	&	25078	&	17046	&   	4966 	& 26719\\
				&		&				&			&			&			&	3595 	& 19874  \\ 
& & & & & & & \vspace{0.5em} \\	
				\hline
& & & & & & & \vspace{0.5em} \\	
MACSJ0416-24	&	3.25	&				&	   		&       		&       		&  	10446 	& 15832 \\ 
& & & & & & & \vspace{-0.5em} \\	
MACSJ0647+70	&	5.18	&	0551850401	&	43342	&	41383	&	29290	&   	3196 	& 19275 \\
				&		&				&			&			&			&	3584 	& 19604 \\
& & & & & & & \vspace{-0.5em} \\	
MACSJ0717+37	&	6.75	&	0672420101	&	41473	&	44476	&	28794	&	4200 	& 58451 \\
				&		&	0672420201	&	54354	&	54393	&	41445	&   			&  \\
				&		&	0672420301	&	42458	&	44462	&	27168	&   			&  \\
& & & & & & & \vspace{-0.5em} \\	
MACSJ1149+22	&	2.32	&				&	    		&        		&      			&   3589 	&20047 \\
				&		&				&			&			&			&	1656 	& 18514 \\
MACSJ2129-07	&	5.00	&				&	      		&      			&        		&   	3199 	& 10847 \vspace*{0.5em} \\
\enddata
\tablenotetext{a}{Soft X-ray absorbing galactic hydrogen column density along line of sight to cluster. }
\tablenotetext{b}{These XMM data were insufficient to define an independent temperature/mass profile.}
\end{deluxetable}

\newpage

\section{X-ray Data Analysis}

Derivations of HSE mass profiles from X-ray data entail a two-step process.  First, one must prepare the data for fitting by selecting a cluster center and dividing the X-ray photon events among a series of concentric annular bins.  Then one must fit the binned two-dimensional data with a projected three-dimensional model, usually a spherically symmetric one.  This section describes the two-step process used to derive the radial profiles presented here.  We use identical procedures to fit both the {\em Chandra} and XMM data, so that the comparisons we make can be as direct as possible.  Our primary fitting tool is the Joint Analysis of Cluster Observations (JACO) code (Mahdavi et al. 2007, 2013), which we describe at the end of the section.  Experienced X-ray observers may find some parts of this section rather basic.  We will attempt to paint a complete picture that allows astronomers outside the X-ray community to see where possible systematic differences between the {\em Chandra} and XMM results can arise.

\subsection{Chandra Data Preparation}

We reprocessed all of the Chandra data identified in Table 1 with CIAO 4.6.1 (released Feb 2014) and  CALDB 4.5.9 (released Nov 2013).  The version and date of the calibration are important to note because this particular calibration revision introduced some important corrections to the soft energy response due to contamination issues.  Time-dependent contamination means that everything from the energy response to the gain has a time dependence.  Calibration changes and updates therefore may introduce differences between today's results and results published in the past.   

In order to remove flares from the data, light curves limited to events between 0.5-7.0 keV were extracted from source-free areas of the detectors.  Any time interval with an event rate $3\sigma$ above the mean event rate was then excluded from the timeline. 
Particularly serious flares were excluded from the timeline using a more conservative $2\sigma$, which effectively removes the ramp-up and ramp-down times near a large flare and the short intervals between major flares. 

Most of the CLASH {\em Chandra} datasets were free of significant flaring. The ``deflare" script in CIAO was used  to generate lists of good time intervals (GTI), which were then applied to remove events acquired during flares from the events files. The majority of these data were taken using ``VFAINT" mode, which allows for a stricter cleaning of background events at the expense of only 1-2\% of the source counts.

Bright point sources were identified and removed using the CIAO detection algorithm {\em wavdetect} together with a map of the Point Spread Function (PSF) size as a function of location on the detector. We filtered regions around these point sources from the event lists. For {\em Chandra} data, point-source excision did not result in a significant loss of cluster event counts or sky area coverage because the
point sources are quite compact.

Backgrounds coming from fainter X-ray point sources, soft X-ray emission from hot gas in our own galaxy, and non-flaring particle events must be accounted for using deep background files.  
Each dataset was therefore matched to a deep background file from a similar observation epoch courtesy of the {\em Chandra} data center  \citep{2007ApJ...661L.117H,2003ApJ...583...70M}). 
These deep-background files, which include both particle and photon backgrounds, were filtered and reprojected to match the target observations. 
As recommended by the Chandra Science Center, we rescaled the reprojected background rates  to match the observation count rates between 10.0-12.0 keV, an energy regime nearly completely dominated by high-energy particle events, not photons. The mechanism we used was to adjust the effective exposure time in the headers of the reprojected background event files. Typically this rescaling affected the effective background exposure time by less than 10-15\%. 

Annular bin boundaries are then selected so that each annular bin contains at least 1500 counts of photon signal from the cluster.  In some cases, these counts come from multiple observations of the same cluster. 
Each annular bin has a minimum radial width at least a few times the PSF width.
Bins that are too narrow can cause the fitting of a 3D model to become unstable, because deprojection and PSF correction have some mutual covariance. 

Once the annular bins are selected, we generate X-ray spectra for each bin from both the image files and the deep background files.  
For {\em Chandra}, spectra were created with the CIAO {\em specextract} script. 
Preparation of X-ray spectra includes the computation of individual weighted redistribution matrix files (RMFs) 
and ancillary response files (ARFs). We extracted spectra binned in energy from 0.5-11.0 keV, with a bin width
of 38 eV.  

At this point the annular spectra are ready for fitting, but with one caveat.  The positional variation of the galactic soft X-ray background is not accounted for by our use of the deep-background files.  Therefore, when fitting any of these datasets, we allow for a possible over- or under-subtraction of the galactic soft X-ray background by fitting the contribution of a  uniform soft thermal X-ray background (fixed at solar abundance), parametrized by its normalization (positive or negative) and temperature. \citep{2007ApJ...664..162M}. Fixing the temperature of this component at 0.5 keV did not affect the outcome.

\subsection{XMM Data Preparation}

Our preparation of the XMM data proceeds along much the same lines.  
We reprocess the XMM Observation Data Files (ODF)  using XMM Science Analysis System (SAS) tasks (version xmmsas\_20120621\_1331) and current calibration files.
The most recently acquired data were processed using version xmmsas\_20131209\_1901-13.5.0
We used data from all 3 spectroscopic-imaging telescope/detector combinations on XMM for the European Photon Imaging Camera (EPIC):
Metal Oxide Semiconductor CCD cameras 1 and 2 (MOS1 and MOS2), and a 3rd, back-illuminated CCD camera called the pn. 
We filtered the event light curves using standard XMM criteria and procedures described in Mahdavi et al. (2013). 

Similarly to the Chandra processing, we exclude bright point sources, using the CIAO task {\em wavdetect} to locate sources on the pn-detector image. 
We always visually inspect the sources identified on this list to make sure not to exclude the cluster itself. 
Occasionally, additional sources are added manually to this list to remove noise spikes near chip boundaries and other sources
of excess counts that {\em wavdetect} missed. 
Usually no more than a few additional excision regions are defined by hand. 
We use relatively source-free regions of the MOS1, MOS2, and pn detectors to assess the particle background between 10-12 keV. To do this, we calculate the 10-12 keV count rate in the blank sky fields provided by the XMM Science Operations Center  \citep{2007A&A...464.1155C}. The ratio of the blank sky 10-12 keV count rate to the observation 10-12 keV count rate is used to normalized the spatially resolved blank sky spectra, which are then subtracted from source spectra prior to any fitting. While this procedure mitigates the particle background, it has the side effect of over- or under-subtracting the astrophysical background (i.e., the contribution of unresolved AGN and the galactic soft X-ray background). To address this side effect, as with the Chandra spectra, we include nuisance background models which are uniform over the field of view of the observations. These backgrounds are modelled as  thermal plasmas with adjustable temperature and, for this work, fixed solar metallicity, with positive or negative normalization specific to each detector. Therefore a total of 5 nuisance parameters for the case of full XMM usage (MOS1+MOS2+pn)  are used to mitigate this residual background.

For XMM data we employ the techniques described in Mahdavi et al (2013) to prepare the spectra for fitting, which includes the subtraction of the ``out-of-time" events from the XMM pn spectra. (Out of time events, which are events that are mis-assigned locations along the readout because they arrive during the short time a detector is read out, are negligible for the MOS and for all but the brightest sources for Chandra.)  As with the {\em Chandra} data, we choose annular bin boundaries so that each bin contains at least 1500 source events and a minimum width of $8^{\prime\prime}$, and include nuisance parameters in the fitting procedure to account for direction-dependent differences in the temperature and normalization of the galactic soft X-ray background.

\subsection{A JACO Primer}

Our primary tool for deriving gas and HSE mass profiles from both {\em Chandra} and XMM data is JACO \citep{2007ApJ...664..162M}, which can provide simultaneous fits to X-ray, S-Z, and weak-lensing data.  
Here we use it to fit only the X-ray data. 
JACO employs parametric models for both the dark-matter density and gas density in the fitting procedure.  
In CLASH-X, the dark-matter profiles are assumed to have a spherically-symmetric NFW form, parametrized by a scale radius $r_{\rm s}$ and a density normalization $\rho_{\rm s}$ at that radius, so that
\begin{equation}
  \rho_{\rm DM}(r) = \frac {4 \rho_{\rm s}} { (r/r_{\rm s}) [ 1+ (r/r_{\rm s}) ]^{2} } \; \; ,
\end{equation}
whereas the gas-density profiles are modeled with a more flexible triple $\beta$-model, with one component multiplied by a radial power law of index $-\alpha$:
\begin{equation}
  n_e(r)  \; =  \; n_{e0} \left(   \frac {r} {r_0} \right)^{-\alpha} \left(  1 +  (\frac {r} {r_{e0}})^2 \right)^{-3\beta_0/2}
                           + \; n_{e1} \left(  1 +  (\frac {r} {r_{e1}})^2 \right)^{-3\beta_1/2}  
                           + \; n_{e2} \left(  1 +  (\frac {r} {r_{e2}})^2 \right)^{-3\beta_2/2}    \; \; .
\end{equation}
If the surface brightness profile is adequately fit by a single $\beta$-model truncated by a power law (i.e., the first term), we set $n_{e1} = n_{e2} = 0$. 
JACO can also allow for a stellar contribution to the total mass-density profile in the form of an Einasto profile, but that feature is not used here.   We do, however, allow for a radial metallicity dependence in the intracluster medium, with a profile
\begin{equation}
  \frac {Z} {Z_\odot} = Z_0 \left( 1 + \frac {r^2} {r_Z^2} \right)^{-3 \beta_Z} \; \; .
\end{equation}
Using this combined parametric model, JACO can compute the projected X-ray spectrum coming from each line of sight through the target cluster by making a strong assumption:  the radial gas-temperature profile is determined by requiring that the gas be in hydrostatic equilibrium in the combined potential well of the dark matter and gas.  In this paper, we will refer to the temperature determined in this way as $T_{\rm JACO}$.

Once JACO has made a parametric cluster model, it generates a synthetic event spectrum for each annular bin by convolving the cluster model with an energy-dependent instrumental PSF and adding a background model.
As described in the data preparation subsections, we subtract the particle spectrum using re-normalized deep background fields, and account for the over- or under-subtraction of the galactic soft X-ray background by fitting it to a parametrized soft thermal component.  These corrections are part of the JACO fitting procedure.

Correction for the instrumental PSF turns out to be critical for the CLASH-X mass-profile comparison, because the cores of galaxy clusters are far brighter than the regions at $r_{2500}$ and $r_{500}$, where we would like to make comparisons with the weak-lensing data.
Proper PSF correction is much more important for XMM than {\em Chandra} because its half energy width (HEW) of its PSF is so much broader ($13-17\arcsec$ for XMM\footnote{\url{http://xmm.esac.esa.int/external/xmm_user_support/documentation/uhb/onaxisxraypsf.html}}  compared to $<1\arcsec$ on axis for 
Chandra\footnote{\url{http://asc.harvard.edu/proposer/POG/html/chap4.html}}, according to the instrument handbooks).
We have compared JACO results with and without PSF correction and find that best-fit masses and temperatures derived from uncorrected models are systematically lower than those for corrected models.  
At least part of this effect comes from the scattering of X-ray photons from the cooler, brighter cores of some clusters to larger radii in the detector, thereby reducing the best-fitting temperatures at those radii.
(See, for example, Maxim Markevitch's white paper on Abell 1835, posted on the Chandra calibration website.\footnote{\url{http://arxiv.org/abs/astro-ph/0205333}})
But those scattering events should be accounted for by the PSF correction procedure.
The results we present in \S\ref{section:T_profiles} suggest that additional XMM PSF corrections might be necessary to bring the XMM and {\em Chandra} results into acceptable agreement.

JACO obtains constraints on all parameters of each cluster model through a Monte Carlo Markov Chain (MCMC) procedure that produces likelihood distributions for all the radial profiles of interest.  The uncertainty ranges on X-ray HSE mass profiles reported in \S\ref{section:mass_profiles} come from this MCMC procedure and represent 68\% confidence intervals.  This ``forward-fitting" procedure fundamentally differs from the ``deprojection" procedure commonly used in X-ray astronomy, which obtains radial temperature and density profiles by sequentially fitting and then subtracting the contribution of each spherical shell, starting from the outermost part of the cluster and finishing in the core \citep[e.g.][]{1999AJ....117.2398M, 2006A&A...459.1007C}.  Instead, JACO fits all shells simultaneously, obtaining stronger constraints on the properties of each shell (which are correlated) at the expense of imposing a parametric model on the overall mass distribution.

We have tested the JACO cluster fits in two different ways.   We have compared them with electron density and pressure profiles from ACCEPT \citep{2009ApJS..182...12C} 
and with results from the deprojection code used in \citet{2010A&A...524A..68E}, and find good agreement.  We have also used JACO to produce non-parametric fits for each cluster to a model in which concentric shells with the same inner and outer radii as the annular bins contain isothermal gas of uniform density.  This is not a ``deprojection" but rather an MCMC fit in which the gas density and gas temperature of each bin are the fitted parameters.  It is not subject to the assumptions of hydrostatic equilibrium and NFW-ness of the mass profile and can therefore be used to check whether those assumptions distort the gas density and temperature profiles derived in the parametric fitting procedure.  The results have greater uncertainties but are statistically consistent with those obtained from the parametric fits.

For both XMM and Chandra spectra, the raw spectrum 
energy bin was 38 eV. We typically limited the spectral fit to 0.7-8.0 keV, but occasionally 
truncated the fits at 7.0 keV. Energy bins were grouped to a minimum of 25 counts per grouped spectral bin. 
We conservatively restrict the fits to radial bins where the signal to total counts ratio exceed 0.25, 
equivalent to the signal to background ratio threshold of 0.3 recommended by \citet{2008A&A...486..359L} to avoid any strong systematics in the background treatment.
All spectra and JACO configuration files are provided in the CLASH public data products site hosted by MAST.{\footnote{
	\url{http://archive.stsci.edu/prepds/clash/} }
Table~\ref{table:xraystats} contains information about the JACO parameterized fit to the full cluster dataset for Chandra and,
where available, XMM data.

\begin{deluxetable}{lrrcrlllrrcrlll}
\tabletypesize{\scriptsize}
\tablecaption{X-ray Fit Statistics\label{table:xraystats}}
\tablewidth{0pt}
\tablecolumns{15}
\tablehead{
\colhead{} & \multicolumn{7}{c}{\em Chandra} & \multicolumn{7}{c}{XMM} \\
\colhead{Name} & \colhead{$N_{\rm Rad}$} & \colhead{$N_{\rm Sp}$} & \colhead{$R_{\rm max}$} & \colhead{DOF} & \colhead{$\chi_\nu^2$} & \colhead{Prob} & \colhead{$S/T$} &
\colhead{$N_{Rad}$} & \colhead{$N_{Sp}$} & {$R_{\rm max}$} & \colhead{DOF} & \colhead{$\chi_\nu^2$} & \colhead{Prob} & \colhead{$S/T$}\\
\colhead{} & \colhead{} & \colhead{} & \colhead{arcmin} & \colhead{} & \colhead{} & \colhead{} & \colhead{ACIS} & \colhead{} &
\colhead{} & \colhead{arcmin} & \colhead{} & \colhead{} & \colhead{} & \colhead{M1/M2/pn}   }
\startdata
&   &   &   &   &    &    &     &     &    &    &    & &    &  \vspace{-0.5em} \\
Abell~209 & 4  & 8  & 3.2   & 616  & 0.973 & 0.673 & 0.80     & 12 & 36 & 7.5  & 1289 & 0.937 & 0.945 & 0.3/0.41/0.27  \\
Abell~383 & 13 & 26 & 2.7     & 1430 & 1.09  & 7E-3  & 0.4  & 12 & 36 & 4.17    & 1553 & 1.075 & 0.019 & 0.35/0.31/0.44 \\
MACS0329  & 5  & 10 & 1.08 & 550  & 1.22  & 3E-4  & 0.9      &  \nodata   & \nodata   & \nodata  &  \nodata &  \nodata & \nodata \\
MACS0429  & 5  & 5  & 0.9   & 230  & 0.98  & 0.57  & 0.93     & 8  & 24 & 4.17    & 3783 & 0.68  & 1.00  & 0.36/0.48/0.28 \\
MACS0744  & 5  & 15 & 2.05 & 633  & 1.06  & 0.13  & 0.5      & 4  & 24 & 1.17 & 783 & 1.09  & 0.04  & 0.72/0.79/0.68\\
&   &   &   &   &    &    &     &     &    &    &    &  &    & \vspace{-0.5em} \\
Abell~611 & 16 & 16 & 3.38 & 1024 & 0.99  & 0.57  & 0.38   & \nodata   & \nodata & \nodata   & \nodata  &  \nodata &  \nodata & \nodata \\
MACS1115  & 10 & 10 & 1.25  & 707  & 1.01  & 0.42  & 0.87     & 8  & 24 & 3.53 & 815  & 1.03  & 0.26  & 0.54/0.61/0.87\\
Abell~1423& 9  & 9  & 5  & 786  & 0.975 & 0.684    & 0.29   & \nodata   & \nodata   & \nodata &  \nodata &  \nodata & \nodata & \nodata \\
Abell~2261& 14 & 14 & 3.8   & 1093 & 0.902 & 0.99  & 0.72   & 16 & 48 & 5.8    & 3104 & 1.046 & 0.034 & 0.26/0.36/0.28 \\
MACS1206   & 10 & 10 & 2.7  & 524  & 0.88  & 0.97  & 0.61     & 6  & 18 & 2.68 & 1149 & 0.98  & 0.7   & 0.44/0.56/0.41 \\
&   &    &   &   &    &    &     &     &    &    &    & &    &  \vspace{-0.5em} \\
CL1226    & 3  & 6  &  0.92  & 186  & 1.02  & 0.41  & 0.84     & 3  & 15 & 5    & 504  & 0.77  & 1.00   & \nodata \\
MACS1311  & 7  & 7  & 0.85 & 305  & 0.9   & 0.89  & 0.85     &  \nodata &  \nodata   & \nodata   & \nodata   &   \nodata &  \nodata & \nodata \\
RXJ1347   & 10 & 30 & 2.54 & 5462 & 1.06  & 6E-4  & 0.7      & 12 & 36 & 3.67 & 2787 & 104   & 0.066  & 0.43/0.48/0.34\\
RXJ1423   & 1  & 11 & 0.47 & 861  & 0.958 & 0.807 & 0.95     &  \nodata   &  \nodata   & \nodata   & \nodata  &  \nodata &  \nodata & \nodata \\
RXJ1532   & 22 & 22 & 2.87 & 2756 & 0.944 & 0.98  & 0.24     & 5  & 15 & 2.83 & 540  & 1.04  & 0.24    & 0.68/0.59/0.54 \\
&   &    &   &   &    &    &     &     &    &    &    &   &    &\vspace{-0.5em} \\
MACS1720  & 6  & 12 & 1.66 & 689  & 1.03  & 0.3   & 0.75     & 8  & 24 & 3.67    & 765 & 1.03  & 0.29    & 0.12/0.31/0.17 \\
MACS1931  & 17 & 17 & 3.69 & 2689 & 1.04  & 0.09  & 0.27     & 11 & 33 & 3.89 & 2405 & 1.03  & 0.10    & 0.26/0.24/0.15 \\
RXJ2129   & 16 & 16 & 4.5  & 1202 & 0.916 & 0.98  & 0.27     & 14 & 42 & 4       & 2253 & 1.002  & 0.46   & 0.49/0.50/0.32 \\
MS2137    & 8  & 16 & 0.82 & 1303 & 0.94  & 0.94  & 0.93     & 6  & 18 & 1.58 & 1880 & 1.11  & 0.0004  & 0.83/0.84/0.73 \\
RXCJ2248   & 18 & 18 & 3.04 & 1320 & 0.925 & 0.975 & 0.72     & 13 & 30 & 3.67 & 2503 & 1.04  & 0.06    & 0.56/0.56/0.45 \\
&   &    &   &   &    &    &     &     &    &    &    & &    &  \vspace{-0.5em} \\
MACS0416  & 3  & 3  & 1.39 & 117  & 1.12  & 0.18  & 0.83     &  \nodata   & \nodata   & \nodata  &  \nodata &  \nodata & \nodata \\
MACS0647  & 3  & 6  & 0.95  & 163  & 0.94  & 0.68  & 0.92     & 5  & 15 & 1.67    & 558  & 1.00  & 0.46    & 0.58/0.71/0.52 \\
MACS0717  & 11 & 11 & 2.2   & 829  & 1.16  & 7E-4  & 0.79     & 7  & 63 & 2    & 3462 & 1.01  & 0.27    & 0.71/0.76/0.59\\
MACS1149  & 3  & 6  & 1.44  & 292  & 0.84  & 0.98  & 0.89     &  \nodata   & \nodata & \nodata  & \nodata  &  \nodata &  \nodata & \nodata \\
MACS2129  & 3  & 6  & 1.58 & 197  & 1.11  & 0.14  & 0.8      &  \nodata  &  \nodata   & \nodata   & \nodata  &  \nodata &  \nodata & \nodata \vspace{-0.5em}  \\
\enddata
\tablecomments{$N_{Rad}$ is the number of radial bins; $N_{Sp}$ is the total number of spectra; $R_{\rm max}$ is outer radius in arcminutes
of the outermost bin. DOF is the number of degrees of freedom (the number of spectral energy bins minus the number of fit parmeters) for the full JACO parameterized fit; $\chi_\nu^2$ is the $\chi^2$ statistic divided by the DOF; Prob is the probability corresponding to the best fit. $S/T$ is the ratio of the total number of source counts $S$ 
divided by the total number of particle and soft X-ray counts $T$ in the 0.7-8.0 keV band in the outermost bin, with the exception of the outermost pn observation for Abell 383, where the ratio is the total source counts to the particle background.}
\end{deluxetable}

\section{Profiles of Gas Properties}
\label{section:gas_profiles}

This section presents the gas properties derived from the JACO fits.  We find very good agreement between the electron-density profiles coming from the {\em Chandra} and XMM datasets, and therefore very good agreement between the gas-mass profiles as well.  However, agreement between the gas-temperature profiles is not as satisfactory.   Temperature agreement between {\em Chandra} and XMM is good in the cluster cores ($\lesssim 200$~kpc) but becomes progressively worse at larger radii, with $T_{\rm JACO}({\rm xmm}) \approx 0.75 \, T_{\rm JACO}({\rm chandra})$ at $r \sim 800$~kpc, indicating either a need for more accurate XMM PSF correction, improved knowledge of the off-axis response, or a background-subtraction issue with either XMM or {\em Chandra}.

\subsection{Gas Density \& Gas Mass}
\label{section:ne_profiles}

The left panel of Figure~\ref{figure:ne_mgas} shows the ratio between XMM-derived gas density and {\em Chandra}-derived gas density as a function of radius for each CLASH-X cluster with both XMM and {\em Chandra} data, along with the unweighted mean, weighted mean, and median ratios at each radius.  Here and throughout the paper, we present {\em geometric} means instead of {\em arithmetic} means for these ratios, so that the means are symmetric with respect to an exchange of the numerator and denominator (i.e. $\langle A/B \rangle = \langle B/A \rangle^{-1}$ for geometric means but not arithmetic means).   Individual clusters show some density-ratio excursions arising from the fitting procedure, because we are fitting parametric electron-density profiles to {\em Chandra} and XMM datasets with different bin spacing.  However, the mean ratios are virtually identical, with a slight trend to smaller densities in the XMM data at large radii.

\begin{figure}
  \begin{minipage}[b]{0.5\linewidth}
    \includegraphics[width=3.5in, trim = 0.6in 0.4in 0.6in 0.5in]{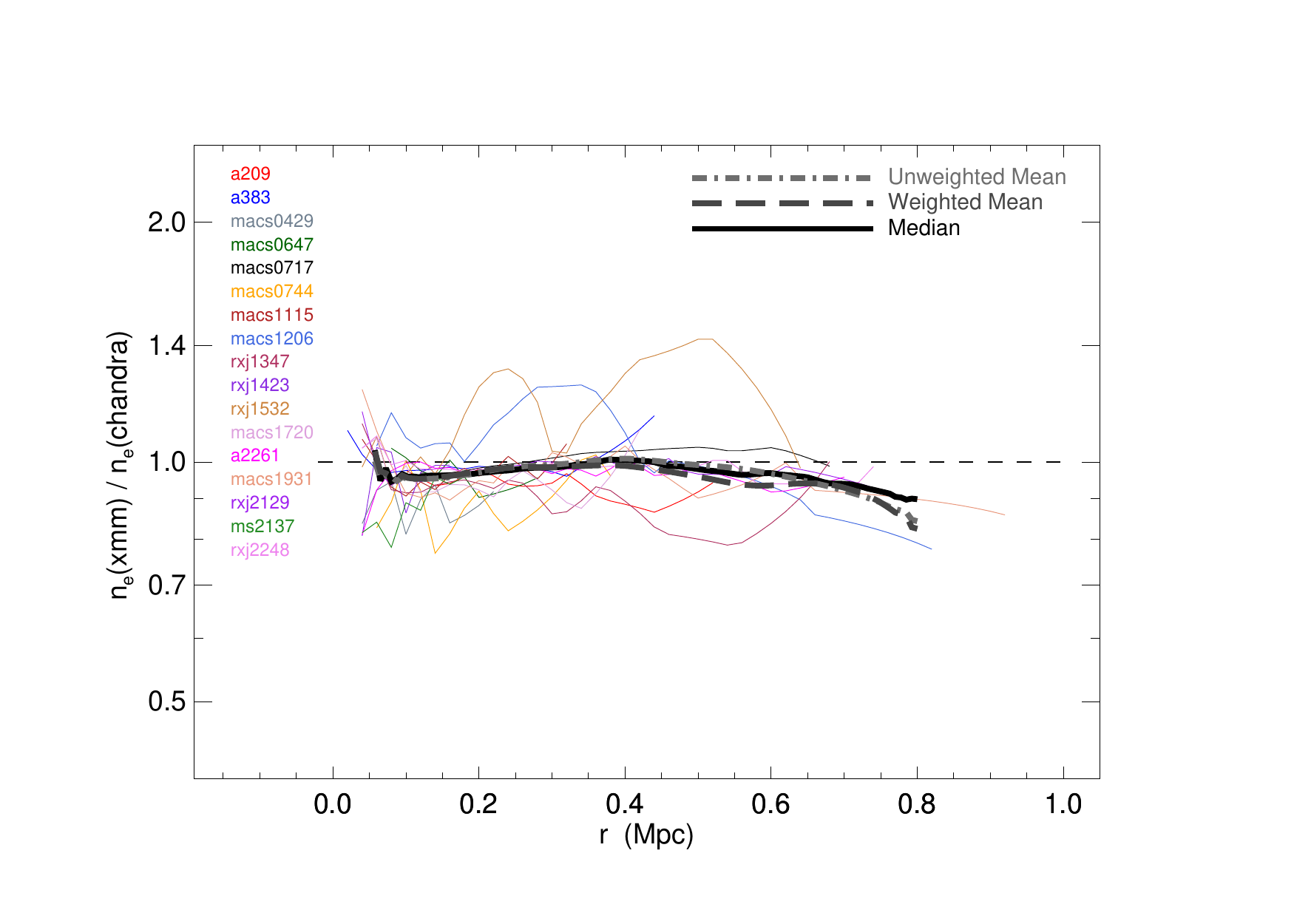}
  \end{minipage}
  \quad
  \begin{minipage}[b]{0.5\linewidth}
    \includegraphics[width=3.5in, trim = 0.6in 0.4in 0.6in 0.5in]{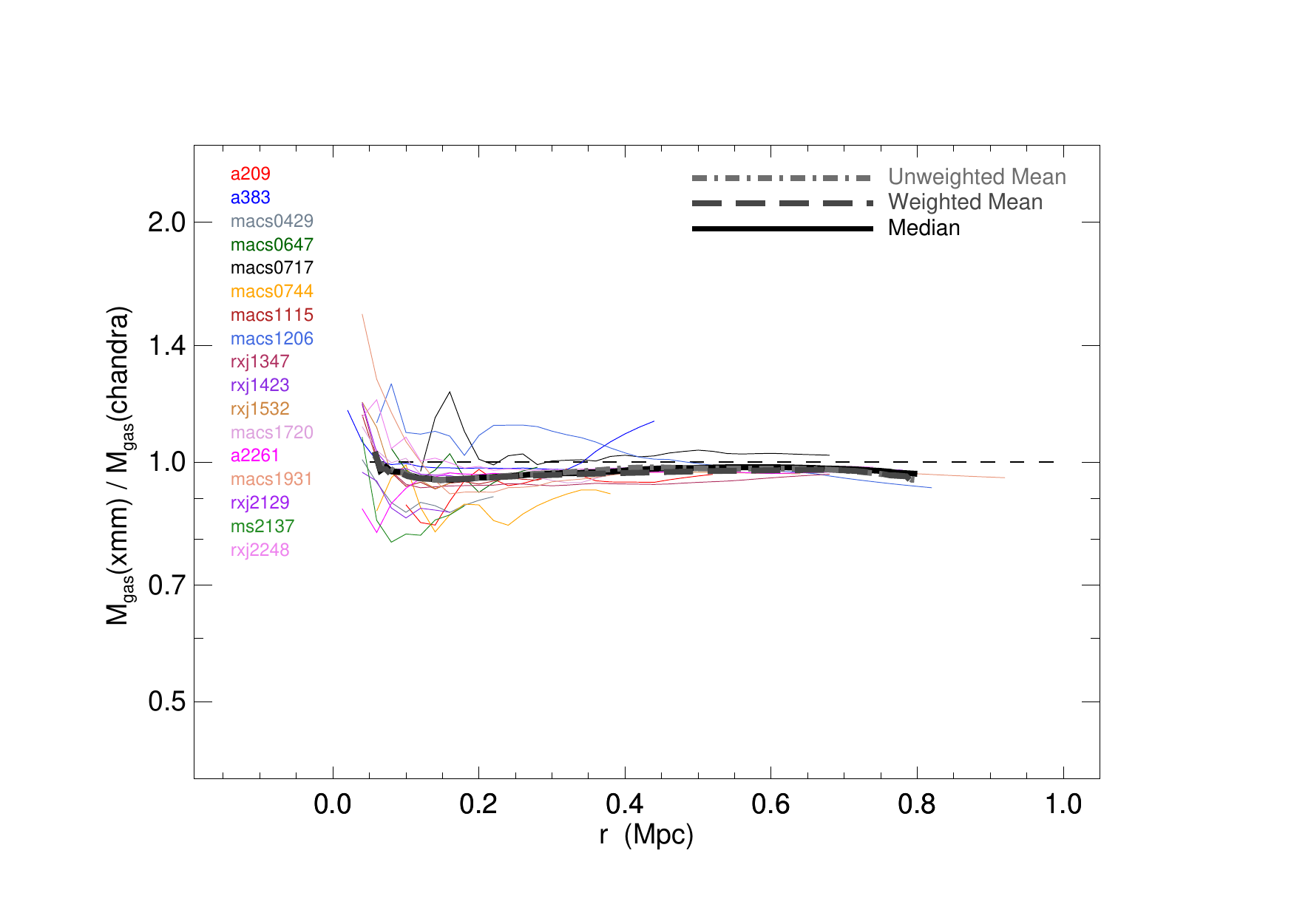} 
  \end{minipage}
  \caption{\footnotesize 
  Average ratios of electron-density $n_e$ (left panel) and enclosed gas-mass $M_{\rm gas}$ (right panel) derived from 
  XMM and {\em Chandra} data using JACO under the assumption of spherical symmetry.  Thick solid lines show 
  the median ratios.  Long-dashed thick lines show weighted means.  Dot-dashed thick lines show unweighted means.  
  Short-dashed lines indicate the locus of equality.  
  Lists at left shows the clusters represented, whose best-fit profile ratios are given by the thin lines.
  \label{figure:ne_mgas}  
  }
\end{figure}

The right panel of Figure~\ref{figure:ne_mgas} shows the ratios of gas mass $M_{\rm gas}(r)$ enclosed within radius $r$ derived by JACO from the XMM and {\em Chandra} data.  Integration of gas density over radius smoothes out the variations seen in the electron-density fits, leading to excellent agreement in the gas-mass profiles outside of 0.2~Mpc with a dispersion of only a few percent.  It is reassuring to see such consistency between the $M_{\rm gas}(r)$ values derived from {\em Chandra} and XMM.

\subsection{Temperature Profiles}
\label{section:T_profiles}

\begin{figure}[t]
\includegraphics[width=7.0in, trim = 0.0in 0.4in 0.0in 0.1in]{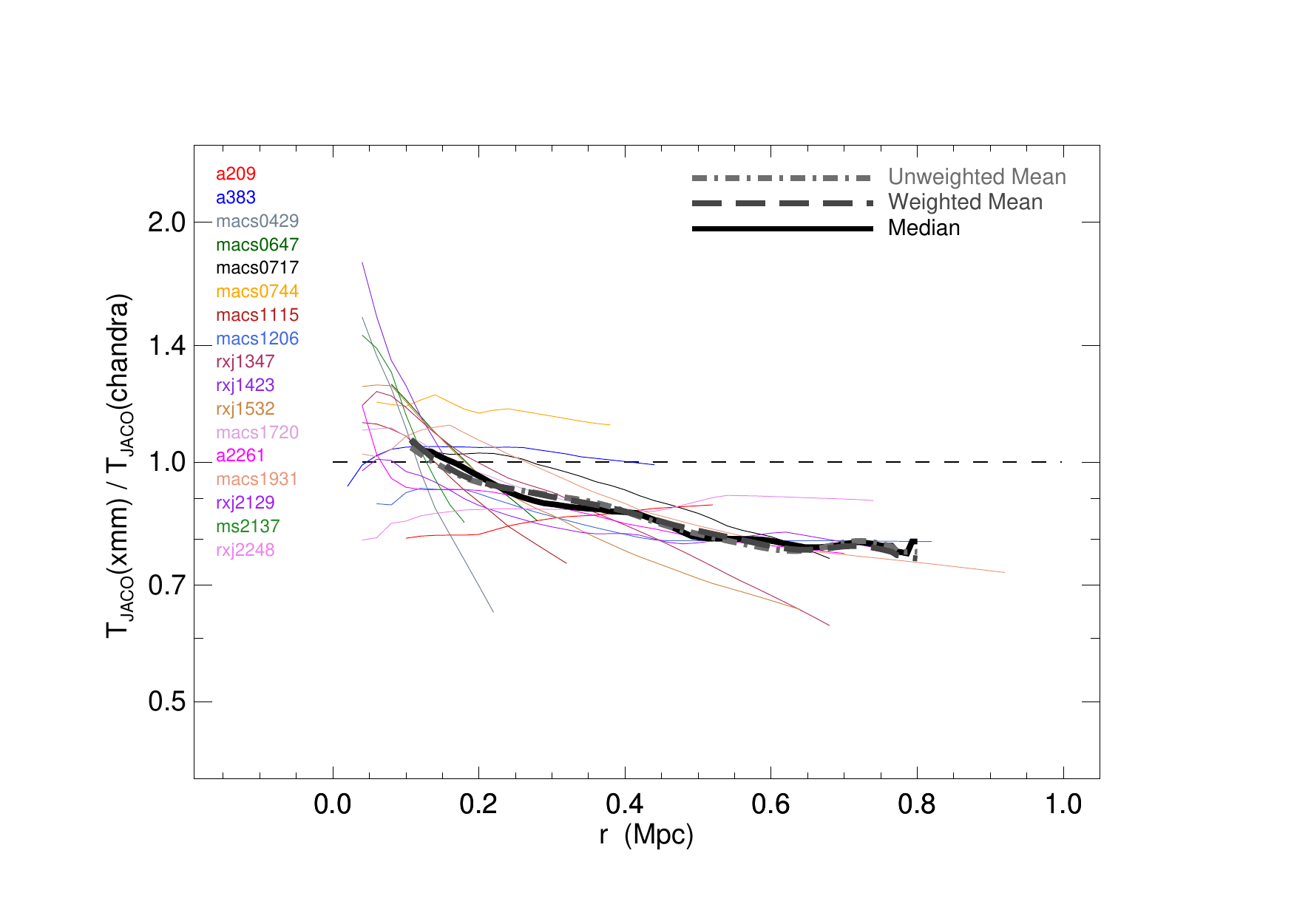} \\
\caption{\footnotesize
  Average ratios of derived gas temperature $T_{\rm JACO}$ from XMM and {\rm Chandra}, as functions of radius.  
  Line styles are the same as in Figure~\ref{figure:ne_mgas}.
  \label{figure:T_JACO_ratio}
  }
\end{figure}

Comparisons of the temperature profiles from {\em Chandra} and XMM are not as comforting.  Figure~\ref{figure:T_JACO_ratio} shows our comparisons.  At $\sim 100$~kpc radii there is no systematic difference between the temperatures JACO measures from {\em Chandra} and XMM data, but the XMM temperature systematically declines relative to the {\em Chandra} temperature as distance from the cluster center increases, reaching a mean ratio $\approx 0.75$ at radii approaching 1~Mpc.

\begin{figure}[t]
\includegraphics[width=7.0in, trim = 0.0in 0.2in 0.0in 0.1in]{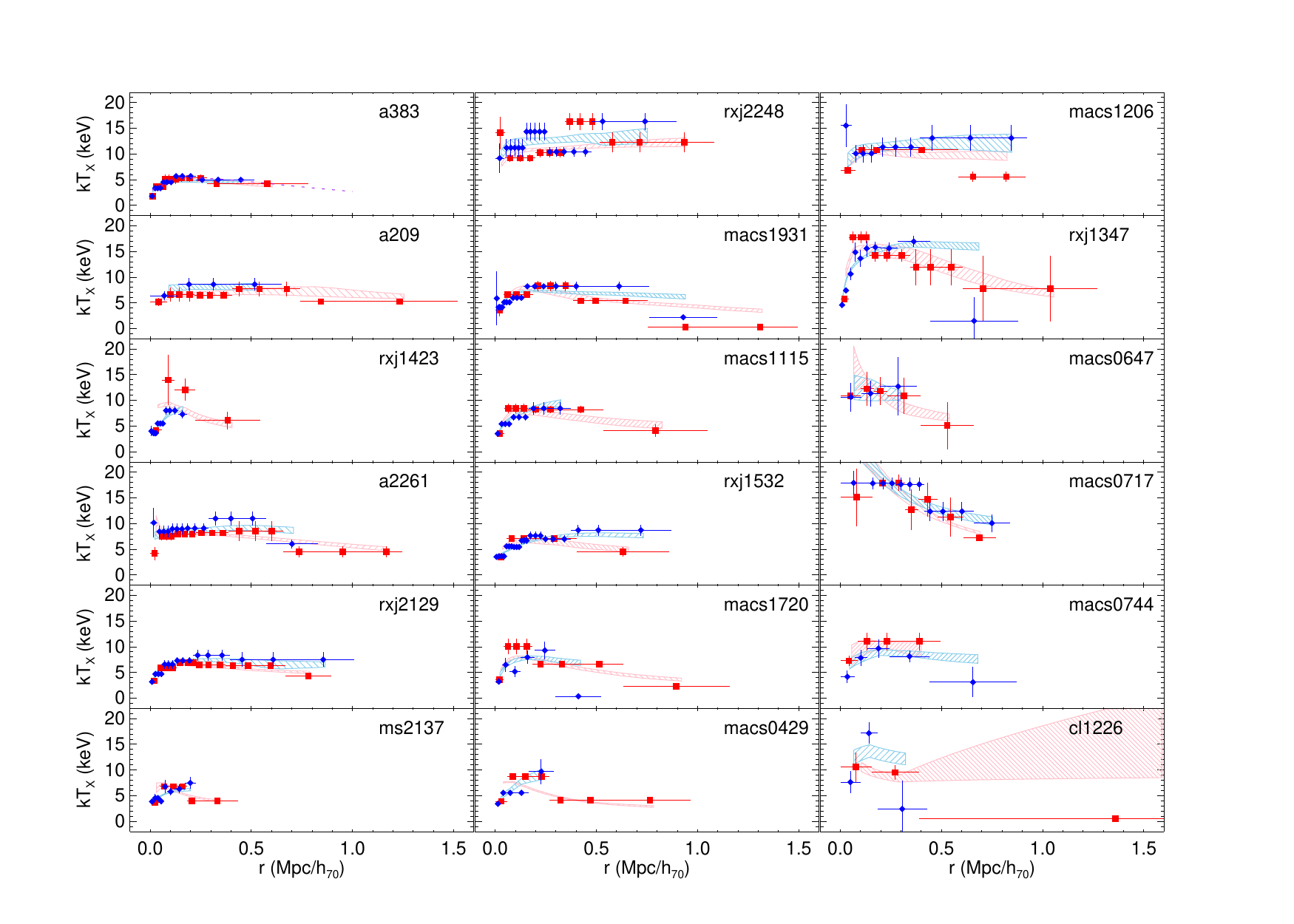} \\
  \caption{\footnotesize 
  Temperature fits to XMM and {\em Chandra} data for CLASH clusters observed with 
  both telescopes.  Red squares show non-parametric fits to XMM data. 
  Blue diamonds show non-parametric fits to {\em Chandra} data.  
  Horizontal error bars on those points represent the radial ranges of the annular regions used to extract the X-ray events. 
  Hatched areas show the temperature profiles corresponding to the best-fitting 
  parametric JACO models, which assume hydrostatic equilibrium and NFW mass profiles.
  The dotted line in the Abell 383 panel shows the best-fitting {\em Chandra} temperature 
  profile for that cluster from  \citet{2006ApJ...640..691V}.  Clusters are ordered by increasing 
  redshift, starting from the top left and proceeding downward.
  \label{figure:T_MCMC}
  }
\end{figure}

Figure~\ref{figure:T_MCMC} supports this finding, because it shows that systematic temperature differences at large radii are not an artifact of the parametric JACO fitting procedure.  Data points in the figure show the non-parametric JACO fits to the binned spectra, which invoke neither the assumption of hydrostatic equilibrium nor the assumption of NFW mass profiles.  They simply represent the uncorrelated bin-by-bin temperatures that best fit the projected spectroscopic data.  And more often than not, the XMM temperatures at $\gtrsim 0.5 $~Mpc are below the {\em Chandra} temperatures, despite the fact that there is no apparent systematic temperature difference at small radii.  We note, as shown in Figure~\ref{figure:T_MCMC}, that the XMM temperature data for cluster CLJ1226+3332 are inadequate for deriving a mass profile.  We have therefore excluded this cluster from the other XMM data comparisons made in this paper.

Cluster temperature discrepancies between XMM and {\em Chandra} have been noticed before.  For example, Mahdavi et al. (2013) found that XMM HSE cluster masses were systematically $\sim 15$\% smaller than {\em Chandra} HSE cluster masses if they made no attempt to correct for a systematic temperature offset.  In order to bring the masses into agreement, they introduced a photon-energy dependent effective-area correction factor of $(E/{\rm keV})^{0.07}$, where $E$ is the photon energy, into the {\em Chandra} data analysis.  An independent comparison by \citet{2014arXiv1404.7130S} shows that XMM temperatures from all three detectors (MOS1, MOS2, and pn) are systematically smaller than {\em Chandra} temperatures by a percentage that increases with cluster temperature and reaches $\sim 20$\% at the $\sim 8$--12~keV temperatures typical of CLASH clusters;
\citet{2010A&A...523A..22N} reach similar conclusions in a comparison of temperatures obtained from  fit to 0.5-2.0 keV spectra compared to those obtained from fitting 2.0-7.0 keV spectra. In that study, the 2.0-7.0 keV results were more similar to Chandra results than those for the 0.5-2.0 keV bandpass.

To our knowledge, our work here shows the first indication that the temperature discrepancy depends on distance from the cluster center.  The fact that our {\em Chandra} and XMM temperatures agree in the core, where photon fluxes are greatest, suggests that miscalibration of either the {\em Chandra} or the XMM effective area on-axis is unlikely to be the main problem.  One example of a systematic that could produce this radial trend is excess large-angle scattering of soft X-ray photons not accounted for in the standard XMM PSF form.  The regions where the temperature differences are greatest in CLASH clusters are typically $\sim 1$~arcmin from the much brighter cluster core.  A relatively small fraction of soft photons scattered from the core into regions $\sim 1$~arcmin away from it could therefore produce a significant decrement in the best-fit temperature. Taking the XMM PSF into account is a challenging endeavor \citep[e.g., ][]{2008A&A...487..431C,2004A&A...423...33P}, and so development using the JACO analysis platform is planned for future work. Alternatively, improper background treatments for either the XMM or {\em Chandra} could result in apparent temperature differences at larger radii, but we consider that hypothesis unlikely because temperature differences persist in regions where the signal-to-background ratio is large. Finally, we cannot rule out the possibility that the XMM and {\em Chandra} responses are compatible at the usual aimpoint, but incur increasing discrepancies in the outer parts of the detectors. This latter possibility is not so far-fetched because most of the calibration is likely to be the best on-axis. Investigation of all of these possibilities is beyond the scope of this paper, but our intention is to explore these issues in a future work.

\section{Mass Profiles}
\label{section:mass_profiles}

This section presents the mass profiles derived for each CLASH cluster from all the data sets available to us at the time of publication.  It presents cluster mass profiles as circular-velocity plots because in that form they are virtually independent of cosmology and therefore allow mass profiles derived assuming different cosmologies to be plotted simultaneously.  The plots show that HSE mass profiles derived from {\em Chandra} observations are often nearly identical to those determined from lensing.  However, the plots also show that many of the XMM mass profiles are systematically tilted to lower masses at larger radii, because of the radial dependence of the temperature offset.  Sections~\ref{section:chandra_lensing} and \ref{section:xmm_lensing} present more detailed comparisons of HSE and lensing mass profiles for {\em Chandra} and XMM, respectively.

\subsection{Circular Velocity \& Enclosed Mass}

In our experience, galaxy-cluster astronomers spend altogether too much time converting masses and mass profiles back and forth among the slightly different $\Lambda$CDM cosmologies they have used to derive those masses.  Much wasted effort could be avoided if they simply provided and compared mass profiles in circular-velocity form.  To see why, consider this form of the hydrostatic equilibrium equation for a spherical mass configuration in which $M_r$ and $\theta_r$ represent the mass enclosed within radius $r$ and angle subtended by that radius at the cluster's distance, respectively, and $\mu m_p$ is the mean mass per gas particle:
\begin{equation}
  v_{\rm circ}^2 (\theta_r) \equiv \frac {G M_r} {r} = \frac {kT(\theta_r)} {\mu m_p} \left| \frac {d \ln P} {d \ln r} \right|_{\theta_r}  \; \; .
\end{equation}
Both of the factors needed to calculate $v_{\rm circ} (\theta_r)$ can be derived from an X-ray observation without invoking a cosmological model, and converting to $v_{\rm circ} (r)$ requires only an angular-size distance.  Likewise, an analogous version of $v_{\rm circ} (\theta_r)$ can be derived from lensing observations, for which the only cosmological dependences stem from the slightly altered distance distribution of the lensed background galaxies. (Conversion to a projected physical radius is necessary to compare mass profiles for clusters at different redshifts, but mass-bias analyses require only comparisons of mass profiles derived for the same cluster using different techniques.)

Aside from cosmological independence, circular velocities have several additional advantages over enclosed masses, and particularly over enclosed masses defined with respect to a spherical-overdensity threshold:
\begin{itemize}

\item Dividing $M_r$ by $r$ removes the lowest-order dependence of $\ln M_r$ on $\ln r$, greatly relieving the compression of dynamic range along the vertical axis of an $M_r(r)$ plot.  Systematic differences among derived mass profiles are then much easier to see by eye, because they are not so highly compressed.

\item  The value of $v_{\rm circ} (r)$ for an NFW profile is nearly constant in the vicinity of the scale radius:  It reaches a maximum value $v_{\rm max}$ at $2.163 \, r_{\rm s}$, remains within 6\% of $v_{\rm max}$ over the interval $1\lesssim r/r_s \lesssim 5$, and stays within 2\% of $v_{\rm max}$ over the interval $1.4 \lesssim r/r_s \lesssim 3.5$.  Circular-velocity measurements are therefore much less subject to aperture-induced covariances than enclosed-mass measurements, in which correlated errors in the determinations of spherical-overdensity radii can induce systematically correlated offsets in all quantities defined with respect to those radii, complicating the task of distinguishing true observational biases from those introduced by the subsequent analysis.

\item  If a dark-matter halo is not changing, then there is no evolution in $r_{\rm s}$ or $v_{\rm max}$.  The same cannot be said for $r_\Delta$, $M_\Delta$, or $c_\Delta$, which continue to change simply because the density threshold used to measure them declines with time.  A halo's value of $v_{\rm max}$ is therefore a more direct indicator of its rarity than $M_\Delta$, because it is independent of redshift.

\item Accurate estimates of $v_{\rm max}$ are often easier to obtain from X-ray observations than accurate estimates of $M_{500}$ or $M_{200}$ because the $v_{\rm circ}$ profile in an NFW potential well with concentration $c_{200} \sim 4$ peaks near $r_{2500}$.  An exact value of $v_{\rm max}$ can be hard to measure because of the lack of curvature in the $v_{\rm circ}(r)$ profile in the vicinity of its peak, but that same feature is advantageous for making accurate estimates, because $v_{\rm max}$ can be reliably estimated from any set of $r$--$M_r$ pairs measured in the $\sim 0.3$--$2$~Mpc range.

\end{itemize}
Because of these considerations, we provide NFW profile fits in the form of $r_{\rm s}$--$v_{\rm max}$ pairs for {\em Chandra} in Table~\ref{table:cxcmasses} and for XMM in Table~\ref{table:xmmmasses}.  The NFW mass profile in terms of those quantities is 
\begin{equation}
  M_r  =  4.625 \, \frac {v_{\rm max}^2 r} {G} \, \left[  \frac{\ln (1 + x)} {x} - \frac {1} {1+x}  \right]  \;  \; ,
\end{equation} 
where $x = r/r_{\rm s}$.  The same tables also provide conventional $M_{2500}$, $r_{2500}$, and $r_{500}$ values, along with the enclosed gas fraction $f_{\rm g,2500} = M_{\rm gas}(r)/M_r$ measured at $r_{2500}$.  Uncertainties in the table correspond to the 68\% confidence regions from the JACO MCMC analysis.

\subsection{Mass Profiles in Circular-Velocity Form}

Figure~\ref{figure:rxj2129} shows examples of mass profiles in the form of $v_{\rm circ}$ curves for CLASH cluster RXJ2129+00.  A dot-dashed (blue) line shows the mass profile derived from {\em Chandra} data by the JACO fitting procedure, assuming hydrostatic equilibrium.  A short-dashed (red) line shows the analogous JACO mass profile derived from {\em XMM} data.  Hatched regions show 68\% confidence intervals from the JACO MCMC chain.  Both profiles are free of cosmological assumptions when viewed as $v_{\rm circ}(\theta_r)$ curves.  The bottom axis shows radial distance (in $h_{70}^{-1}$~Mpc) from the cluster center, which is uncertain by $\lesssim 1$\% at $z \lesssim 0.5$ within a flat $\Lambda$CDM cosmology in which the value of $\Omega_{\rm M}$ is known to $\lesssim 10$\%.  Cosmological uncertainties in the positions of the light-gray diagonal lines showing enclosed mass are therefore also $\lesssim 1$\%.

\begin{deluxetable}{lcccccccccccc}
\tabletypesize{\scriptsize}
\tablecaption{Chandra HSE Masses\label{table:cxcmasses}}
\tablewidth{0pt}
\tablehead{
\colhead{Name} & \colhead{$v_{\rm max}$} & \colhead{$\sigma_{v_{\rm max}}$} &\colhead{$r_{\rm s}$} & \colhead{$\sigma_{r_{\rm s}}$} & \colhead{$r_{2500}$} & \colhead{$\sigma_{r_{\rm 2500}}$} &\colhead{$M_{2500}$} & \colhead{$\sigma_{M_{\rm 2500}}$}  &\colhead{$f_{\rm g,2500}$}& \colhead{$\sigma_{f_{\rm g,2500}}$}& \colhead{$r_{500}$}  & \colhead{$\sigma_{r_{\rm 500}}$} \\
\colhead{} & \colhead{km s$^{-1}$} & \colhead{} &\colhead{$h_{70}^{-1}$ Mpc} & \colhead{} & \colhead{$h_{70}^{-1}$ Mpc} &\colhead{} & \colhead{$10^{14} h_{70}^{-1}$ M$_\odot$} & \colhead{}  & \colhead{$h_{70}^{-3/2}$} & \colhead{} & \colhead{$h_{70}^{-1}$ Mpc}  & \colhead{}}
\startdata
	&   	&	&     &	&	&	&	&	&	&	&	& \vspace{-0.5em}  \\
Abell 209	&1743  	&126	&0.745	&0.653	&0.47	&0.014	&2.49	&0.36	&0.11	&0.005	&1.31	&0.16 \\
Abell 383	&1184	&29 	&0.22	&0.017	&0.436	&0.007	&1.42	&0.07	&0.107	&0.003	&0.94	&0.021 \\
MACS0329-02	&1485	&80 	&0.36	&0.10	&0.46	&0.017	&2.24	&0.24	&0.117	&0.007	&1.05	&0.056 \\
MACS0429-02	&1496	&171	&0.30	&0.10	&0.486	&0.034	&2.49	&0.57	&0.105	&0.01	&1.07	&0.105 \\
MACS0744+39	&1631	&84 	&0.43	&0.16	&0.425	&0.015	&2.34	&0.24	&0.128	&0.007	&1.01&0.07 \\
	&   	&	&     &	&	&	&	&	&	&	&	& \vspace{-0.5em}  \\
Abell 611	&1678	&92 	&0.57	&0.17	&0.552	&0.019	&3.20	&0.35	&0.088	&0.0044	&1.30	&0.09 \\
MACS1115+01	&1664	&106	&0.45	&0.097	&0.546	&0.023	&3.30	&0.42	&0.113	&0.0077	&1.25	&0.091 \\
Abell 1423	&1297	&61 	&0.275	&0.051	&0.47	&0.014	&1.82	&0.17	&0.094	&0.0037	&1.03	&0.042 \\
MACS1206-08	&2002	&148	&0.71	&0.507	&0.587	&0.03	&4.59	&0.68	&0.122	&0.0077	&1.43	&0.16 \\
CLJ1226+33*	&4975&		    &3.89	&6.38	&0.705	&0.05	&13.6	&2.90	&0.038	&0.007	&2.30	&0.36 \\
	&   	&	&     &	&	&	&	&	&	&	&	& \vspace{-0.5em}  \\
MACS1311-03	&1390	&116	&0.32	&0.094	&0.42	&0.022	&1.80	&0.30	&0.108	&0.01	&0.95	&0.08 \\
RXJ1347  	&2318	&57 	&0.403	&0.03	&0.735	&0.012	&9.14	&0.45	&0.104	&0.003	&1.60	&0.035 \\
MACS1423+24	&1579	&146 	&0.287	&0.047	&0.472	&0.026	&2.70	&0.50	&0.099	&0.0075	&1.04	&0.07 \\
MACS1532+30	&1629	&41 	&0.46	&0.042	&0.525	&0.009	&3.00	&0.15	&0.119	&0.004	&1.22	&0.033 \\
MACS1720+35	&1467	&89 	&0.273	&0.05	&0.483	&0.018	&2.40	&0.29	&0.114	&0.006	&1.05	&0.055 \\
	&   	&	&     &	&	&	&	&	&	&	&	& \vspace{-0.5em}  \\
Abell2261	&1571	&56 	&0.304	&0.059	&0.567	&0.013	&3.24	&0.23	&0.1096	&0.004	&1.23	&0.045 \\
MACS1931-26	&1522	&33 	&0.279	&0.02	&0.511	&0.008	&2.74	&0.12	&0.133	&0.0035	&1.11	&0.023 \\
RXJ2129+00	&1488	&70 	&0.34	&0.043	&0.529	&0.017	&2.67	&0.25	&0.105	&0.004	&1.17	&0.048 \\
MS2137-2353	&1312	&44 	&0.164	&0.016	&0.449	&0.01	&1.78	&0.12	&0.1205	&0.0048	&0.93	&0.027 \\
RXJ2248-44	&2272	&125	&0.828	&0.326	&0.706	&0.025	&7.19	&0.79	&0.1226	&0.0067	&1.71	&0.14 \\
	&   	&	&     &	&	&	&	&	&	&	&	& \vspace{-0.5em}  \\ \hline
	&   	&	&     &	&	&	&	&	&	&	&	& \vspace{-0.5em}  \\
MACS0416-24	&2000	&800	&\nodata &$<8$	&0.6	&0.1	&3.8	&1.4	&0.094	&0.016	&2.1	&0.35 \\
MACS0647+70 &2600   &640	&1.26   &3.58   &0.62   &0.07   &6.5    &3.2    &0.008  &0.013  &1.67   &0.55 \\
MACS0717+37	&2300	&110	&0.96	&0.38	&0.59	&0.02	&5.4	&0.5	&0.124	&0.004	&1.52	&0.13 \\
MACS1149+22	&1700	&200	&\nodata &$<4$	&0.49	&0.04	&3.1	&0.8	&0.123	&0.008	&1.10	&0.28 \\
MACS2129-07	&2000	&400	&\nodata &$<5$	&0.56	&0.06	&4.7	&1.7	&0.009	&0.015	&1.35	&0.62  \vspace{-0.5em} \\
\enddata
\end{deluxetable}

\begin{deluxetable}{lcccccccccccc}
\tabletypesize{\scriptsize}
\tablecaption{XMM HSE Masses and Maximum Circular Velocities \label{table:xmmmasses}}
\tablewidth{0pt}
\tablehead{
\colhead{Name} & \colhead{$v_{\rm max}$} & \colhead{$\sigma_{v_{\rm max}}$} &\colhead{$r_{\rm s}$} & \colhead{$\sigma_{r_{\rm s}}$} & \colhead{$r_{2500}$} & \colhead{$\sigma_{r_{\rm 2500}}$} &\colhead{$M_{2500}$} & \colhead{$\sigma_{M_{\rm 2500}}$}  &\colhead{$f_{\rm g,2500}$}& \colhead{$\sigma_{f_{\rm g,2500}}$}& \colhead{$r_{500}$}  & \colhead{$\sigma_{r_{\rm 500}}$} \\
\colhead{} & \colhead{km s$^{-1}$} & \colhead{} &\colhead{$h_{70}^{-1}$ Mpc} & \colhead{} & \colhead{$h_{70}^{-1}$ Mpc} &\colhead{} & \colhead{$10^{14} h_{70}^{-1}$ M$_\odot$} & \colhead{}  & \colhead{$h_{70}^{-3/2}$} & \colhead{} & \colhead{$h_{70}^{-1}$ Mpc}  & \colhead{}}
\startdata
	&   	&	&     &	&	&	&	&	&	&	&	& \vspace{-0.5em}  \\
Abell 209	&1506	&48	&0.81	&0.17	&0.46	&0.01	&1.73	&0.11	&0.125	&0.003	&1.2	&0.03 \\
Abell 383	&1243	&27	&0.24	&0.02	&0.45	&0.007	&1.61	&0.07	&0.095	&0.003	&0.98	&0.02 \\
MACS0429-02	&1347	&25	&0.059	&0.003	&0.4	&0.004	&1.36	&0.05	&0.135	&0.002	&0.77	&0.01 \\
MACS0744+39	&1568	& 65 & 0.14 &0.02 &0.43	&0.013	& 2.4 & 0.2 	&0.115	&0.005	&0.88	&0.03 \\
	&   	&	&     &	&	&	&	&	&	&	&	& \vspace{-0.5em}  \\
MACS1115+01	&1546	&51	&0.25	&0.03	&0.52	&0.011	&2.89	&0.19	&0.12	&0.005	&1.1	&0.03 \\
MACS1206-08	&1974	&70	&0.64 	&0.15	&0.59	&0.013	&4.66	&0.32	&0.12	&0.006	&1.4	&0.1 \\
CLJ1226+33*	&3667	&220&0.2&0.2 	&0.87	&0.03	&25  	&3	    &0.095	&0.003	&3.7	&0.3 \\
RXJ1347 	&2096	&105&0.19	&0.02	&0.65	&0.022	&6.4	&0.64	&0.122	&0.005	&1.33	&0.05 \\
	&   	&	&     &	&	&	&	&	&	&	&	& \vspace{-0.5em}  \\
MACS1423+24 &1534	&68	&0.06	&0.01	&0.413	&0.012	&1.8	&0.2	&0.11	&0.01	&0.80	&0.03 \\							
MACS1532+30	&1423	&61	&0.18	&0.03	&0.475	&0.014	&2.22	&0.19	&0.138	&0.008	&1.00	&0.04 \\
MACS1720+35 &1409	&37	&0.16	&0.02	&0.461	&0.008	&2.1	&0.11	&0.118	&0.003	&0.95	&0.02 \\
Abell 2261	&1491	&24	&0.28	&0.03	&0.539	&0.006	&2.78	&0.09	&0.116	&0.002	&1.15	&0.03 \\
	&   	&	&     &	&	&	&	&	&	&	&	& \vspace{-0.5em}  \\
MACS1931-26	&1489	&35	&0.23	&0.03	&0.502	&0.007	&2.59	&0.12	&0.133	&0.003	&1.07	&0.03 \\
RXJ2129+00	&1331	&30 &0.21   &0.02	&0.478	&0.008	&1.97	&0.09	&0.122	&0.003	&1.01	&0.02 \\
MS2137-2353	&1297	&36	&0.06	&0.01	&0.401	&0.007	&1.27	&0.07	&0.133	&0.004	&0.78	&0.02 \\
RXJ2248-44	&2265	&54	&1.02	&0.18	&0.668	&0.011	&6.08	&0.29	&0.133	&0.004	&1.70	&0.07  \\
	&   	&	&     &	&	&	&	&	&	&	&	& \vspace{-0.5em}  \\ \hline
	&   	&	&     &	&	&	&	&	&	&	&	& \vspace{-0.5em}  \\
MACS0647+70	&1753	&144&0.22	&0.09	&0.52	&0.03	&3.7	&0.6	&0.105	&0.007	&1.1	&0.1 \\
MACS0717+37	&2065	&40	&0.6 	&0.15	&0.59	&0.007	&5.2 	&0.2 	&0.129	&0.002	&1.4	&0.05 \vspace{-0.5em} \\
\enddata
\end{deluxetable}

\begin{figure}
 \includegraphics[width=7.0in, trim = 0.1in 0.4in 0.6in 0.0in]{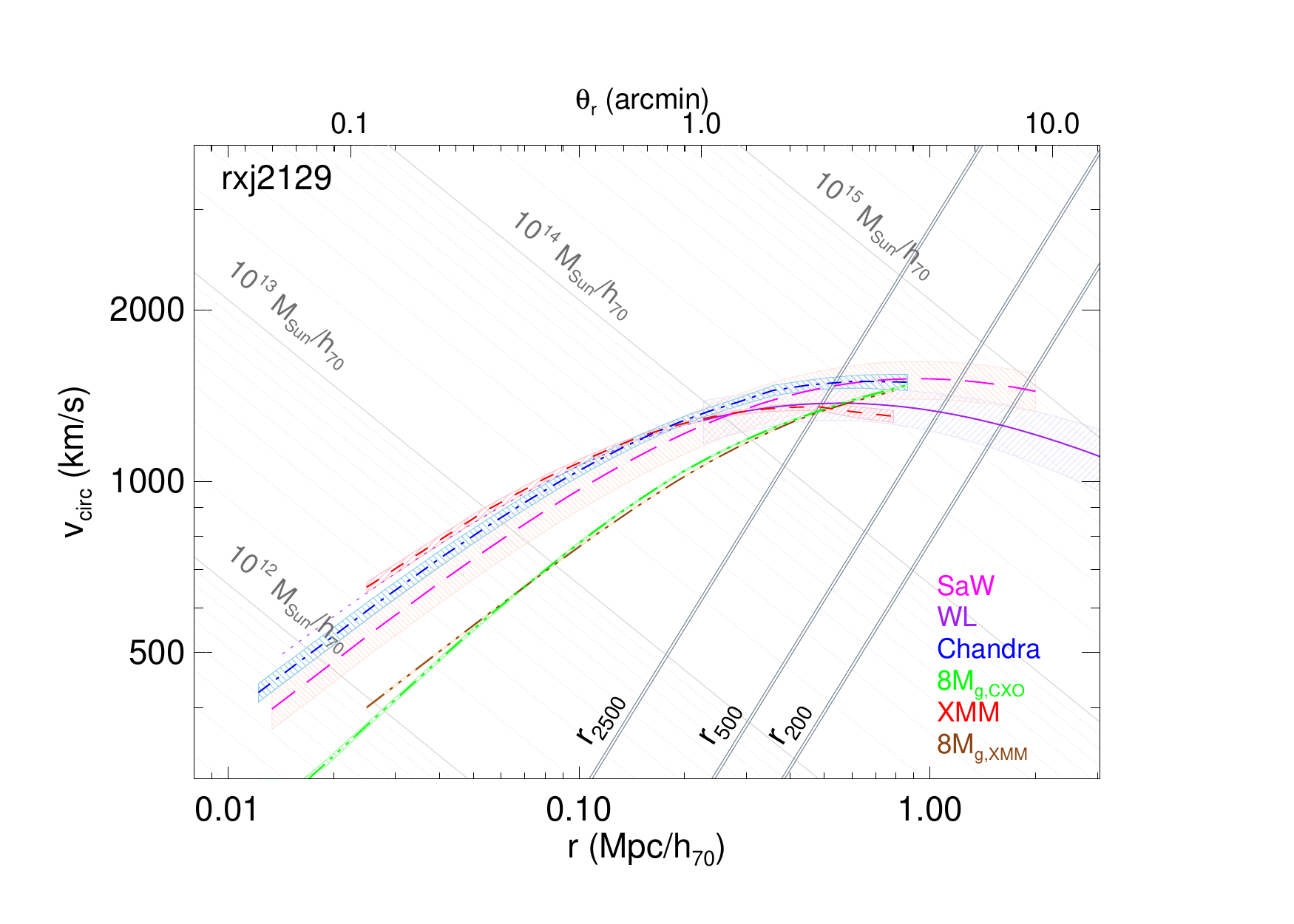} \\
 \caption{ \footnotesize
 Observed 3D deprojected mass profiles for RXJ2129, presented in $v_{\rm circ}(r)$ form and derived from the SaWLens 
 (long-dashed line), WL (solid line), {\em Chandra}/JACO (dot-dashed line), and XMM/JACO (short-dashed line) 
 data sets.  Also shown are  gas-mass profiles, multiplied by a factor of 8, from the parametric JACO analyses of 
 {\em Chandra} (light triple dot-dashed line) and XMM (dark triple dot-dashed line) data.  
 In each case, hatched areas show the 68\% confidence regions.  The hatched area for the weak-lensing (WL) data 
 does not extend inside of 1~arcmin, because the WL fit is less reliable there, but a dotted line shows the inward 
 extrapolation of that fit. \vspace*{1.0em}
 \label{figure:rxj2129}
 }
\end{figure}

Another set of darker gray diagonal lines shows the loci of the spherical-overdensity radii $r_{2500}$, $r_{500}$, and $r_{200}$.  Two lines for each $r_{\Delta}$ indicate the cosmological uncertainty corresponding to $\Omega_{\rm M} = 0.3 \pm 0.03$ at the redshift of the cluster.  Intersections between these lines and the $v_{\rm circ}(r)$ curves give the values of $M_{\Delta}$ determined by each method.  Notice that systematic differences in $M_{\Delta}$, measured along lines of constant $r_{\Delta}$, are larger than the systematic differences in $M_r$, measured in the vertical direction, by an amount that depends on the local slopes of the $v_{\rm circ}(r)$ curves.  Measurements of the average mass-bias factor $\langle b_X \rangle$ therefore depend on whether it is measured at fixed physical radius or at fixed overdensity.  This is one manifestation of the effect sometimes called aperture-induced covariance, in which uncertainties in mass measurement techniques produce correlated uncertainties in all quantities measured with respect to a spherical-overdensity radius.

Solid (purple) and long-dashed (magenta) lines show the mass profiles inferred from the CLASH weak-lensing (WL) and combined strong-and-weak lensing (SaW) data.  The weak-lensing profile at angular radii $< 1$~arcmin, where the weak-lensing fit becomes less statistically secure, is shown with a dotted line.  There is an additional systematic uncertainty of $\pm8\%$ in the overall mass calibration of CLASH weak lensing (Umetsu et al. 2014). Such a systematic would serve to move all profiles in the analysis up or down.
When shown as $v_{\rm circ}(\theta)$ curves, the only sensitivity to cosmological assumptions in these lensing mass profiles comes from slight shifts in the distance distribution of the lensed background galaxies.

\begin{figure}
 \includegraphics[width=7.0in, trim = 1.2in 0.0in 1.5in -0.15in]{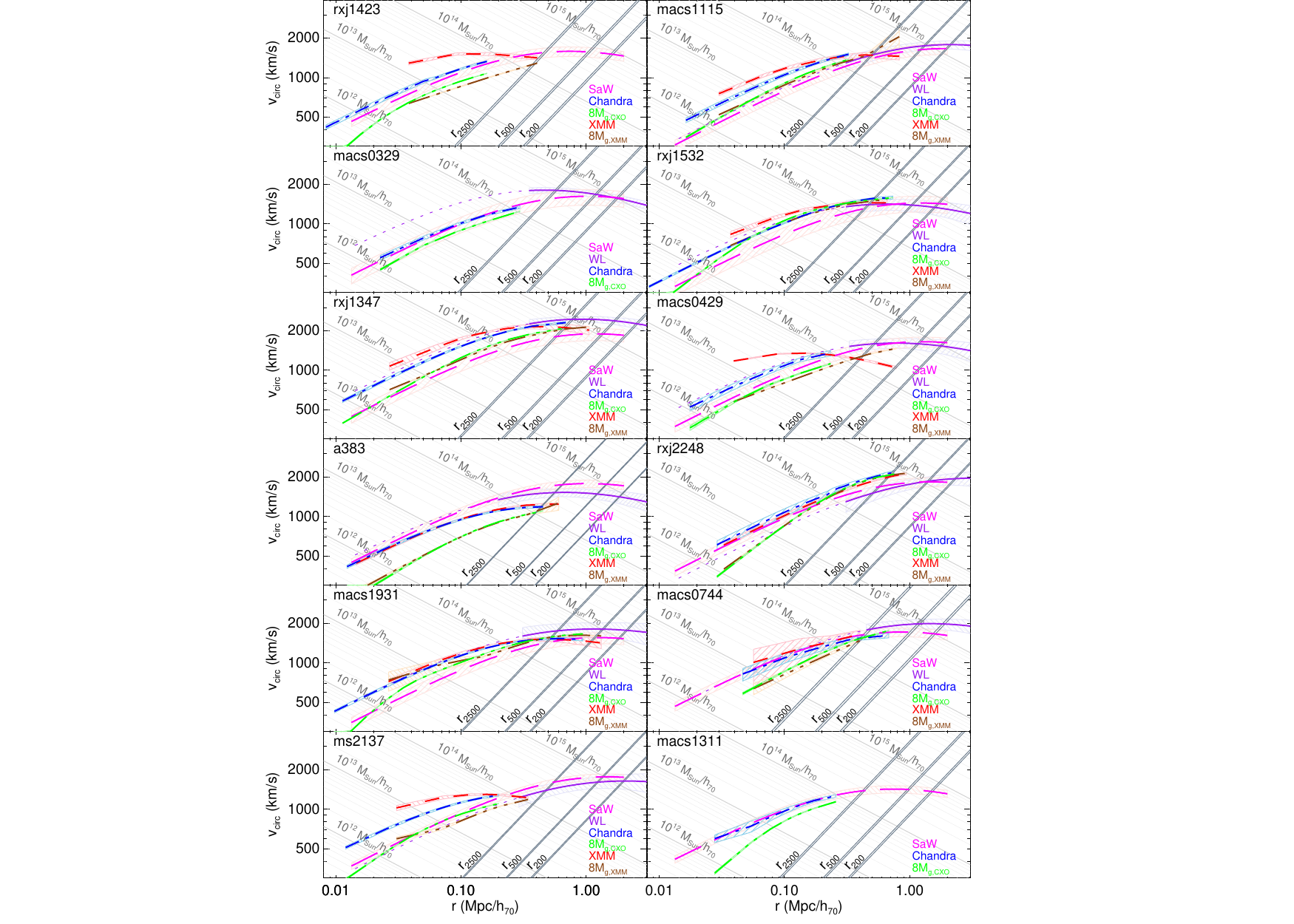} \\
 \caption{ \footnotesize
 Circular velocity plots for the 12 CLASH clusters with $K_0 < 50 \, {\rm keV \, cm^2}$.  
 Line coding and conventions are the same as in Figure~\ref{figure:rxj2129}.
 Core entropy ($K_0$) increases from $10.2 \, {\rm keV \, cm^2}$ to $14.7 \, {\rm keV \, cm^2}$ proceeding down the left column and from $14.8 \, {\rm keV \, cm^2}$ to $47.4 \, {\rm keV \, cm^2}$ proceeding down the right column.  
 \label{figure:Vcirc_2x6_Set1}
 }
\end{figure}

\begin{figure}
 \includegraphics[width=7.0in, trim = 1.2in 0.0in 1.5in -0.15in]{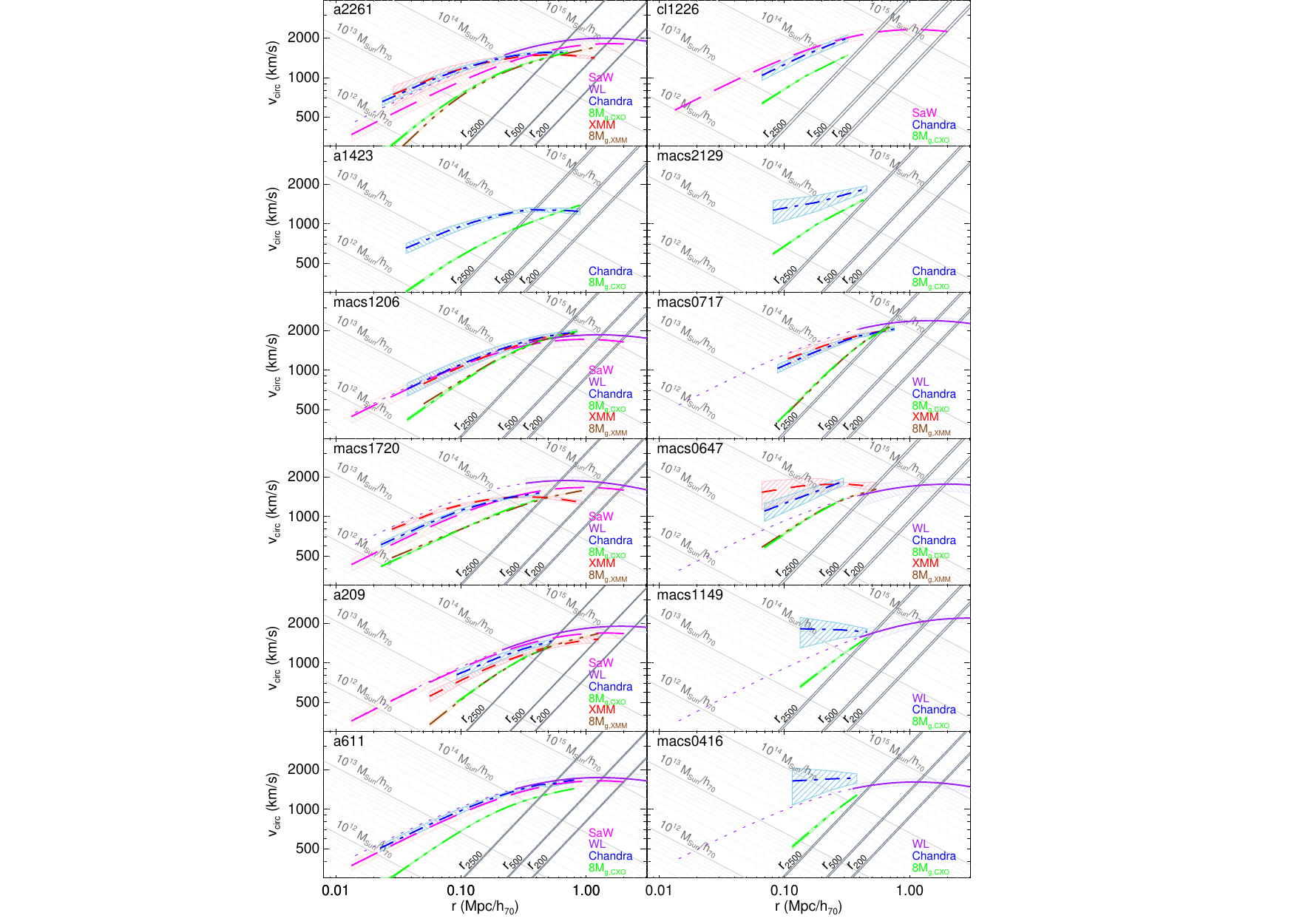} \\
 \caption{ \footnotesize
 Circular velocity plots for the 12 CLASH clusters with $K_0 > 50 \, {\rm keV \, cm^2}$.  
 Line coding and conventions are the same as in Figure~\ref{figure:rxj2129}.
 Core entropy ($K_0$) increases from $61.1 \, {\rm keV \, cm^2}$ to $125 \, {\rm keV \, cm^2}$ proceeding down the left column and from $166 \, {\rm keV \, cm^2}$ to $400 \, {\rm keV \, cm^2}$ proceeding down the right column.  The last five clusters in the right column are the high-magnification subset.
 \label{figure:Vcirc_2x6_Set2}
 }
\end{figure}

Figure~\ref{figure:rxj2129} also shows radial gas-mass profiles inferred by JACO from {\em Chandra} (light green triple-dot-dashed line) and XMM (dark brown triple-dot-dashed line).  We have multiplied these profiles by a factor of 8, because the ratio of gas mass to total mass typically found at $\gtrsim r_{2500}$ in massive relaxed clusters is $\approx 1/8$ \citep{2003ApJ...590...15V,2004MNRAS.353..457A,2002MNRAS.334L..11A,2008MNRAS.387.1179M,2009A&A...501...61E}. We therefore expect $8 M_{\rm gas}$ to be a reasonably accurate mass-profile estimator outside the cores of clusters and can test this expectation with the CLASH cluster observations (see \S\ref{section:mgas_lensing}).

Figures~\ref{figure:Vcirc_2x6_Set1} and \ref{figure:Vcirc_2x6_Set2} provide mass profiles in $v_{\rm circ}(r)$ form for the other 24 CLASH clusters, in which several patterns can be seen:
\begin{itemize}

\item {\em Chandra} HSE profiles generally have shapes similar to the lensing-mass profiles, with offsets to both greater and lesser masses.

\item {\em XMM} HSE profiles tend to be tilted to higher masses at small radii and lower masses at larger radii, compared to the {\em Chandra} HSE and lensing profiles.  This tilt is a direct consequence of the temperature trend shown in Figure~\ref{figure:T_JACO_ratio}.

\item The $8 M_{\rm gas}$ profiles from both {\em Chandra} and XMM are quite consistent with the other mass measures at $\gtrsim r_{2500}$, even in the dramatically unrelaxed high-magnification subset. Therefore, gas mass appears to be a robust mass proxy.

\item X-ray HSE masses for the unrelaxed, high-magnification subset have larger uncertainties than in the more centrally concentrated clusters but are reasonably similar to the weak-lensing masses in radial regions where the mass profiles overlap.

\end{itemize}

The following three sections focus more closely on comparisons of {\em Chandra} HSE mass, XMM HSE mass, and $M_{\rm gas}$ with the lensing-mass profiles presented
in Umetsu et al. (2014) and Merten et al. (2014).

\section{Chandra-Lensing Comparison} 
\label{section:chandra_lensing}

The CLASH cluster sample does not provide a definitive calibration of the ratio of {\em Chandra} HSE mass to lensing mass, because it is not a statistically complete sample, defined with respect to a particular survey threshold, such as an X-ray flux limit, S-Z signal-to-noise ratio, or optical richness.  It is also rather small for the purpose, currently containing only 18 clusters with mass profiles from both {\em Chandra} and {\em Subaru} weak-lensing data.  And even among that set, only 11 have regions of significant radial overlap.  Also, we note that by requiring the mass profiles to fit an NFW profile, we are not yet accounting for the effect on the gravitational potential of the BCG, so interpretation of X-ray and SaWLens results inside 50-100 kpc should be made with caution.  On the other hand, the overall quality and large radial range of the CLASH lensing data make it an excellent sample for identifying systematic differences between these mass-profile measurement techniques.

\begin{figure}
  \begin{minipage}[b]{0.5\linewidth}
    \includegraphics[width=3.5in, trim = 0.6in 0.4in 0.6in 0.5in]{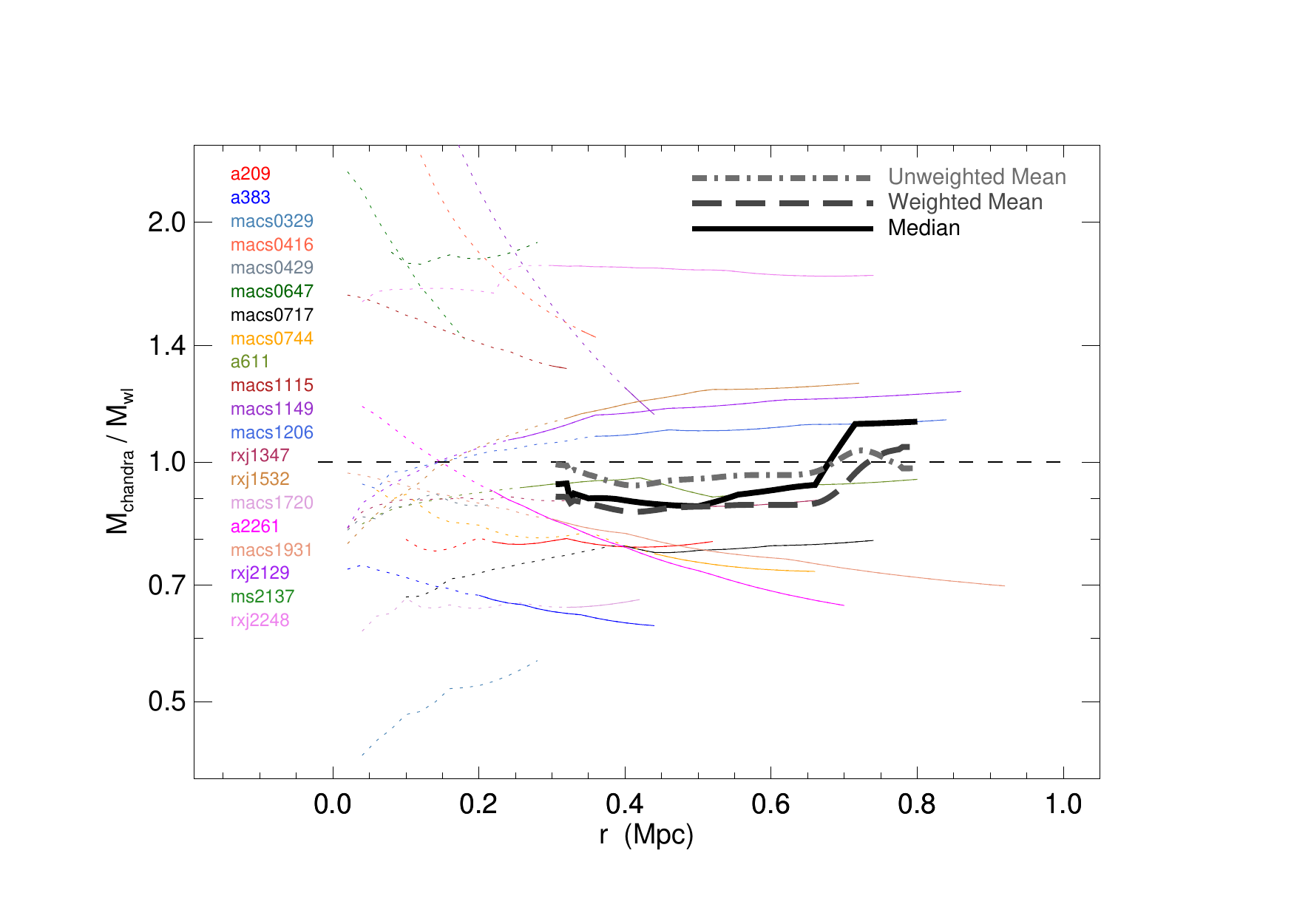}
  \end{minipage}
  \quad
  \begin{minipage}[b]{0.5\linewidth}
    \includegraphics[width=3.5in, trim = 0.6in 0.4in 0.6in 0.5in]{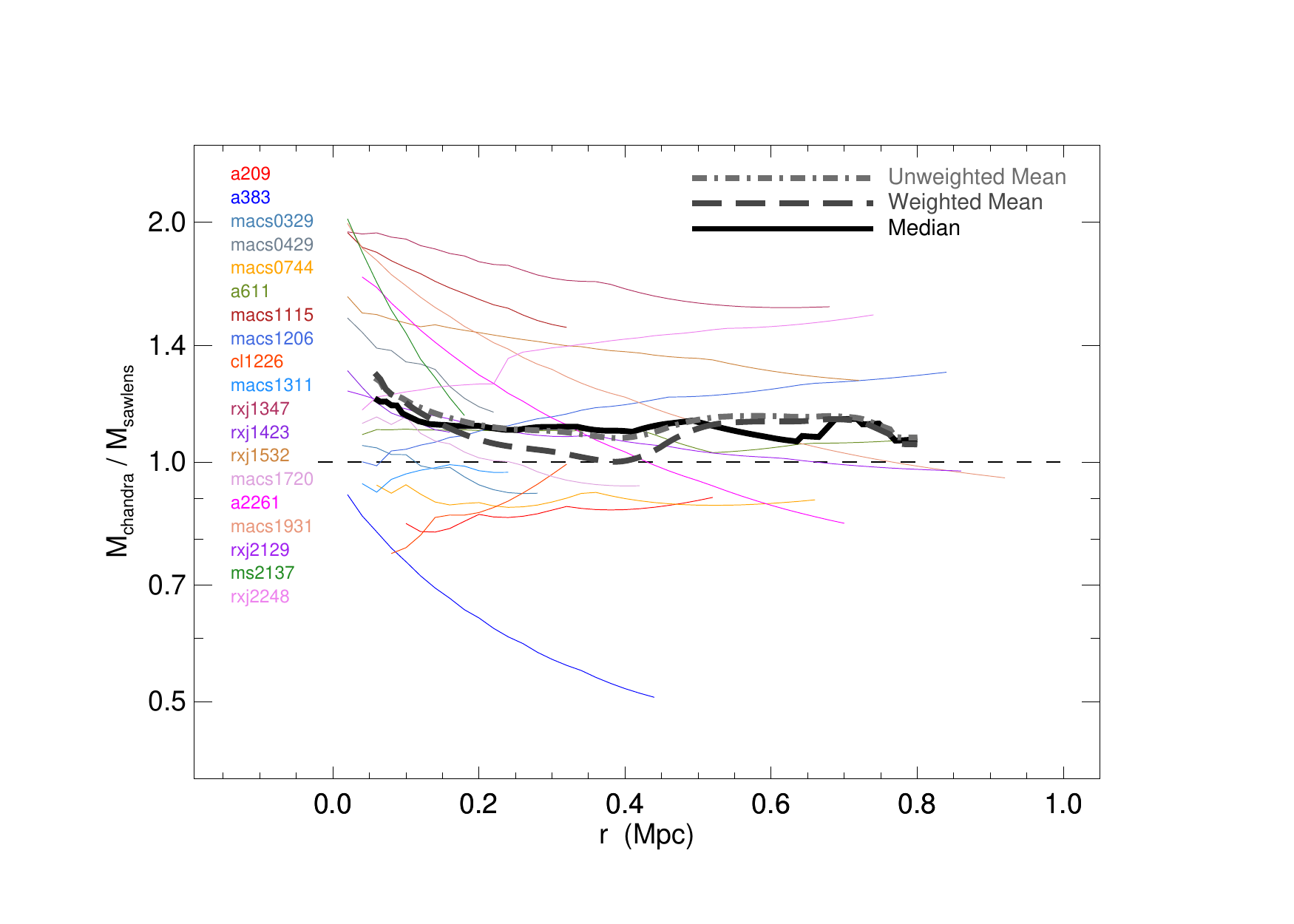} 
  \end{minipage}
  \caption{\footnotesize 
  Average ratios of JACO HSE mass profiles from {\em Chandra} data to the CLASH weak-lensing (left panel) and 
  strong+weak lensing (right panel) profiles.  
  Thick solid lines show the median ratios.  
  Long-dashed thick lines show weighted means.  
  Dot-dashed thick lines show unweighted means.  
  Short-dashed lines indicate the locus of equality.  
  Lists at left shows the clusters represented, whose best-fit profile ratios are given by the thin lines.
  Dotted extensions to those lines in the left panel show extrapolations inside of 1~arcminute, where the best-fitting
  NFW models to the WL data are not well constrained and are not used to compute means or medians.
  \vspace*{1.0em}
  \label{figure:massrat_chandra}  
  }
\end{figure}

Figure~\ref{figure:massrat_chandra} shows the ratios of {\em Chandra} HSE mass to lensing mass we obtain from the CLASH clusters as a function of physical radius.  The left panel compares {\em Chandra} with lensing-mass profiles from {\em Subaru} weak lensing alone.  The right panel compares {\em Chandra} with lensing-mass profiles from the SaWLens combination of {\em Subaru} weak-lensing data with {\em Hubble} data.  For clarity we suppress the uncertainty ranges for individual clusters, which can be inferred from the plots in \S\ref{section:mass_profiles}.  Dotted lines show extensions of the weak-lensing mass-profile fits within 1~arcmin, where the fits become less reliable.  

In the left panel, there is no significant radial trend in the mass-profile ratios, indicating that one should expect similar NFW scale radii and concentrations from both mass-measurement methods.  The scatter is quite significant, with a standard deviation of 0.07 about the unweighted mean at $0.5$~Mpc, but this is within the range expected from projected large-scale structure in the weak-lensing measurements \citep[e.g.,][]{2011ApJ...740...25B}.

The right panel shows no strong radial trend at $\gtrsim 0.2$~Mpc in the comparison to SaWLens profiles, but there is a systematic mass excess in the {\em Chandra} profiles within that radius, rising to $\sim20$\% inside $\sim 100$~kpc.  An astrophysical origin for this excess is implausible, since it would imply that the potential well at small radii is insufficient for balancing the pressure of the hot gas at small radii.  One potential algorithmic origin is inaccuracy of the NFW mass model used to do the fitting, which does not account for the distribution of the stellar mass of the massive galaxies at the centers of many of these clusters.  As one moves inward from $\sim 100$~kpc, the stellar mass fraction becomes increasingly important, and it dominates at $\lesssim 10$~kpc.  

Both strong-lensing and X-ray HSE techniques are sensitive to the stellar mass of the BCG, but stellar mass is likely to have a greater effect on the best-fitting X-ray mass profile in a relaxed galaxy cluster because the prominence of the central X-ray surface-brightness peak causes the innermost parts of an X-ray observation to have greater statistical weight. This effect will be greatest in clusters with the sharpest central peaks in X-ray brightness, which are the ones with the lowest values of core entropy ($K_0$).  One can see signs of the effect in Figures~\ref{figure:Vcirc_2x6_Set1} and \ref{figure:Vcirc_2x6_Set2}.  Virtually all of the clusters in which the {\em Chandra} HSE mass significantly exceeds the SaWLens mass belong to the low-entropy subset in Figure~\ref{figure:Vcirc_2x6_Set1}, with the greatest differences in RXJ1347, MACS1931, MS2137,  MACS1115, RXJ1532, and MACS0429, all of which have $K_0 < 20 \, {\rm keV \, cm^2}$. Among the higher core-entropy subset in Figure~\ref{figure:Vcirc_2x6_Set2}, only the lowest-entropy example shows such a {\em Chandra} mass excess at small radii:  Abell~2261 ($K_0 = 61.1\, {\rm keV \, cm^2}$).

\begin{deluxetable}{llcccccccc}
\tabletypesize{\scriptsize}
\tablecaption{ Average Mass Bias in terms of Mass Ratio: $f = 1 - b = M_1/M_2$ \label{table:mass_bias} }
\tablewidth{7in}
\tablehead{
\colhead{ $M_1$ } & 
\colhead{ $M_2$ } & 
\colhead{ Weighting } & 
\colhead{ $r$ ($h_{70}^{-1}$ Mpc) } & 
\colhead{ $N_{\rm cl}$} & 
\colhead{ $\langle f \rangle$ } &
\colhead{ $\sigma_{\langle f \rangle}$ } &
\colhead{ $\sigma_{\ln f}$ } &
\colhead{ $\chi_\nu^2$ } &
\colhead{ $\sigma_{\ln f}^{\rm intrinsic}$ } 
}
\startdata
$M_{\rm chandra}$  &  $M_{\rm wl}$             &   unweighted   &   0.5  &  11  &  0.95  &   0.07  &  0.27 & 2.52  &    0.19   \\ 
$M_{\rm chandra}$  &  $M_{\rm wl}$             &     weighted      &   0.5  &  11  &  0.88  &   0.07  &  0.22 & 2.22  &    0.16   \\ 
 & & & & & & & & \vspace{-0.5em} \\
$M_{\rm chandra}$  &  $M_{\rm SaWLens}$  &   unweighted   &  0.5   &  10  &  1.13  &   0.07  &   0.20  &  1.17  &    0.08  \\ 
$M_{\rm chandra}$  &  $M_{\rm SaWLens}$  &     weighted      &  0.5   &  10  &  1.11  &   0.07  &   0.19  &  1.16  &    0.07  \\ 
 & & & & & & & & \vspace{-0.5em} \\
$M_{\rm xmm}$        &  $M_{\rm wl}$             &   unweighted   &   0.5  &  14  &  0.84  &   0.06  &  0.30 & 2.91  &    0.21   \\ 
$M_{\rm xmm}$        &  $M_{\rm wl}$             &     weighted      &   0.5  &  14  &  0.76  &   0.05  &  0.23 & 2.46  &    0.18   \\ 
 & & & & & & & & \vspace{-0.5em} \\
$M_{\rm xmm}$        &  $M_{\rm SaWLens}$  &   unweighted   &  0.5   &  13  &  0.89  &   0.10  &  0.33 & 3.64  &    0.30  \\ 
$M_{\rm xmm}$       &  $M_{\rm SaWLens}$  &     weighted      &  0.5   &  13  &  0.82  &   0.08  &  0.33  & 3.45  &    0.28  \\ 
 & & & & & & & & \vspace{-0.5em} \\
$M_{\rm xmm}$       &  $M_{\rm wl}$             &   unweighted   &   0.8  &    8  &  0.77  &   0.09  &  0.36 &  5.93  &    0.32   \\ 
$M_{\rm xmm}$       &  $M_{\rm wl}$             &     weighted      &   0.8  &    8  &  0.69  &   0.08  &  0.30 &  5.28  &    0.28   \\ 
 & & & & & & & & \vspace{-0.5em} \\
$M_{\rm xmm}$       &  $M_{\rm SaWLens}$  &   unweighted   &  0.8   &    8  &  0.95  &   0.10  &    0.29   &  2.15  &    0.22  \\ 
$M_{\rm xmm}$       &  $M_{\rm SaWLens}$  &     weighted      &  0.8   &    8  &  0.96  &   0.11  &    0.28   &  2.14  &    0.21  \\ 
 & & & & & & & & \vspace{-0.5em} \\
$M_{\rm xmm}$       &  $M_{\rm wl}$             &   unweighted   &   1.1  &    3  &  0.60  &   0.09  &  0.13 &  1.30  &    0.08   \\ 
$M_{\rm xmm}$       &  $M_{\rm wl}$             &     weighted      &   1.1  &    3  &  0.56  &   0.05  &  0.12 &  0.99  &    \nodata   \\ 
 & & & & & & & & \vspace{-0.5em} \\
$M_{\rm xmm}$       &  $M_{\rm SaWLens}$  &   unweighted   &  1.1   &    3  &  0.76  &   0.07  &   0.16   &  0.52  &  \nodata  \\ 
$M_{\rm xmm}$       &  $M_{\rm SaWLens}$  &     weighted      &  1.1   &    3  &  0.75  &   0.07  &    0.10   &  0.52  & \nodata  \\ 
 & & & & & & & & \vspace{-0.5em} \\ \hline
 & & & & & & & & \vspace{-0.5em} \\
$M_{\rm chandra}$       &  $M_{\rm wl}$      &   unweighted   &   $r_{500}$ & 20 & 0.91 & 0.12  &  0.76  & 3.83  & 0.50  \\ 
$M_{\rm chandra}$       &  $M_{\rm wl}$      &     weighted   &   $r_{500}$  & 20 & 0.78  & 0.10  & 0.54  & 3.53 & 0.46 \\ 
 & & & & & & & & \vspace{-0.5em} \\
$M_{\rm chandra}$       &  $M_{\rm SaWLens}$  & unweighted & $r_{500}$ & 19 & 0.95 & 0.15 & 0.66 & 11.3  & 0.65 \\ 
$M_{\rm chandra}$       &  $M_{\rm SaWLens}$  & weighted & $r_{500}$ & 19 & 0.69 & 0.09 & 0.58 & 8.7 & 0.54 \\ 
 & & & & & & & & \vspace{-0.5em} \\
$M_{\rm xmm}$       &  $M_{\rm wl}$      &   unweighted   &   $r_{500}$  & 16 &  0.59  &  0.07  &  0.52  &  3.44  &  0.38   \\ 
$M_{\rm xmm}$       &  $M_{\rm wl}$      &     weighted      &   $r_{500}$  & 16 &  0.56  &  0.06  &  0.43  &  3.39  & 0.36   \\ 
 & & & & & & & & \vspace{-0.5em} \\ 
$M_{\rm xmm}$       &  $M_{\rm SaWLens}$  &  unweighted  &   $r_{500}$  & 15 &  0.61 & 0.10 & 0.61  & 15.7 & 0.60  \\ 
$M_{\rm xmm}$       &  $M_{\rm SaWLens}$   &     weighted   &   $r_{500}$  & 15 &  0.53 & 0.08 & 0.58 & 14.7  & 0.56  \\ 
 & & & & & & & & \vspace{-0.5em} \\ \hline
 & & & & & & & & \vspace{-0.5em} \\ 
$M_{\rm SaWLens}$  &  $M_{\rm wl}$         &   unweighted   &   0.5  &  16  &  0.89  &   0.05  &  0.23  &  0.90  &  \nodata   \\ 
$M_{\rm SaWLens}$  &  $M_{\rm wl}$         &     weighted      &   0.5  &  16  &  0.91  &   0.05  &  0.23  &  0.89  &  \nodata   \\ 
 & & & & & & & & \vspace{-0.5em} \\
$M_{\rm SaWLens}$  &  $M_{\rm wl}$            &   unweighted   &  1.0   &  16  &  0.92  &   0.05  &  0.22  &   1.00  &  \nodata  \\ 
$M_{\rm SaWLens}$  &  $M_{\rm wl}$            &     weighted      &  1.0   &  16  &  0.93  &   0.05  &  0.23  &   1.00  &  \nodata  \\  
 & & & & & & & & \vspace{-0.5em} \\
$M_{\rm SaWLens}$  &  $M_{\rm wl}$            &   unweighted   &  $r_{500}$    &  16  &  0.90  &   0.06  &  0.29  &   1.12  &  0.09  \\ 
$M_{\rm SaWLens}$  &  $M_{\rm wl}$            &     weighted      &  $r_{500}$    &  16  &  0.89  &   0.07  &  0.27  &   1.12  &  0.09  \\  
 & & & & & & & & \vspace{-0.5em} \\ \hline
 & & & & & & & & \vspace{-0.5em} \\ 
$M_{\rm gas}(Chandra)$  &  $M_{\rm wl}$   &   unweighted   &   0.5  &  11  &  0.107  &   0.008  &  0.30   &  4.01  &    0.24   \\ 
$M_{\rm gas}(Chandra)$  &  $M_{\rm wl}$   &    weighted      &   0.5  &  11  &  0.094  &   0.007  &  0.23   &  2.98  &    0.19   \\ 
 & & & & & & & & \vspace{-0.5em} \\
$M_{\rm gas}(Chandra)$  &  $M_{\rm SaWLens}$  & unweighted   &   0.5  & 10  & 0.128  & 0.010  & 0.24 & 1.90 &  0.17   \\ 
$M_{\rm gas}(Chandra)$  &  $M_{\rm SaWLens}$  &    weighted   &   0.5  &  10 &  0.124  &  0.010 & 0.23 &  1.85 &  0.16   \\ 
 & & & & & & & & \vspace{-0.5em} \\ 
$M_{\rm gas}({\rm XMM})$    &  $M_{\rm wl}$   &   unweighted   &   0.5  &  14  &  0.106  &   0.008  & 0.30  & 3.80  &  0.25   \\ 
$M_{\rm gas}({\rm XMM})$    &  $M_{\rm wl}$   &   weighted    &   0.5  &  15  &  0.093  &   0.006  &  0.25  & 2.95  &    0.21   \\ 
 & & & & & & & & \vspace{-0.5em} \\
$M_{\rm gas}({\rm XMM})$    &  $M_{\rm SaWLens}$  & unweighted  & 0.5 & 12  &  0.120  & 0.010 & 0.31 & 2.71 &  0.25   \\ 
$M_{\rm gas}({\rm XMM})$   &  $M_{\rm SaWLens}$  &   weighted  &  0.5  & 12  &  0.117  & 0.011 &  0.35 & 2.70  & 0.24   \\ 
 & & & & & & & & \vspace{-0.5em} \\ 
$M_{\rm gas}(Chandra)$  &  $M_{\rm wl}$   &   unweighted   &   0.8  &  4  &  0.118  &   0.015  &  0.25   &  2.75  &    0.21   \\ 
$M_{\rm gas}(Chandra)$  &  $M_{\rm wl}$   &    weighted      &   0.8  &  4  &  0.121  &   0.016  &  0.23   &  2.71  &    0.18   \\ 
 & & & & & & & & \vspace{-0.5em} \\
$M_{\rm gas}(Chandra)$  &  $M_{\rm SaWLens}$  & unweighted   &   0.8  &  4  & 0.128  & 0.014  & 0.23 &  1.71  &  0.15   \\ 
$M_{\rm gas}(Chandra)$  &  $M_{\rm SaWLens}$  &    weighted   &   0.8  &  4 &  0.122  &  0.013 & 0.19 &  1.60 &    0.12   \\ 
 & & & & & & & & \vspace{-0.5em} \\ 
$M_{\rm gas}({\rm XMM})$    &  $M_{\rm wl}$   &   unweighted   &   0.8  &   8  &  0.114  & 0.013  &  0.36  &  6.59  &   0.29   \\ 
$M_{\rm gas}({\rm XMM})$     &  $M_{\rm wl}$   &     weighted    &   0.8  &  8  &  0.095  &   0.008  &  0.23 &  3.97  &   0.20   \\ 
 & & & & & & & & \vspace{-0.5em} \\
$M_{\rm gas}({\rm XMM})$    &  $M_{\rm SaWLens}$  & unweighted & 0.8  & 8  &  0.141  &  0.012  & 0.27 &  1.45  & 0.13  \\ 
$M_{\rm gas}({\rm XMM})$     &  $M_{\rm SaWLens}$  &   weighted  &   0.8  &   8  &  0.135  & 0.011  & 0.22 & 1.41  & 0.12 \\  
 & & & & & & & & \vspace{-0.5em} \\
$M_{\rm gas}({\rm XMM})$    &  $M_{\rm wl}$  & unweighted &   1.1  &   3  &  0.095  &   0.003  & 0.06 &  0.32  & \nodata  \\ 
$M_{\rm gas}({\rm XMM})$     &  $M_{\rm wl}$  &   weighted    &   1.1  &   3  &  0.093  &   0.010  & 0.22 &  1.43  & 0.12 \\   
 & & & & & & & & \vspace{-0.5em} \\
$M_{\rm gas}({\rm XMM})$    &  $M_{\rm SaWLens}$  & unweighted & 1.1 &  3 & 0.120 & 0.007 & 0.11 & 0.28 & \nodata  \\ 
$M_{\rm gas}({\rm XMM})$     &  $M_{\rm SaWLens}$  &   weighted & 1.1 &  3  & 0.121 & 0.007 & 0.08 &  0.28  & \nodata \\  & & & & & & & & \vspace{-0.5em} 
\enddata
\tablecomments{There is an additional systematic uncertainty of 8\% in the overall mass calibration for the weak lensing profiles for CLASH.} 
\end{deluxetable}
 
Table~\ref{table:mass_bias} gives the average mass biases we find in CLASH-X in terms of mass-ratio factors $\langle f \rangle \equiv  1 - \langle b \rangle$.  We have chosen to compute {\em Chandra}--lensing mass bias at 0.5~Mpc because a comparison at that radius includes almost all of the CLASH clusters with significant regions of radial overlap between our {\em Chandra} and weak-lensing mass profiles.  It is also is in the vicinity of $r_{2500}$ for all of these clusters.  We have chosen not to compare mass biases at $r_{2500}$ in order to avoid introducing additional systematic effects stemming from uncertainties in the mass measurements from which $r_{2500}$ is inferred.  The table lists both a weighted and an unweighted mean mass-ratio factor, $\langle f \rangle$, which is a geometric mean computed at radius $r$, along with the uncertainty $\sigma_{\langle f \rangle}$ in each mean, the fractional standard deviation $\sigma_{\ln f}$ about the mean, a reduced $\chi_\nu^2$ value based on the formal uncertainties in the cluster mass measurements, and the intrinsic fractional dispersion $\sigma_{\ln f}^{\rm intrinsic}$ that remains, in order to obtain a reduced chi-squared of unity after accounting for the formal statistical uncertainties.

Values of the mass-ratio factors for both the {\em Chandra}--WL and {\em Chandra}--SaWLens comparisons at 0.5~Mpc are within 15\% of unity, but are on opposite sides of unity.  The weighted mean mass bias for {\em Chandra}--WL mass bias is  $\langle b \rangle = 0.12 \pm 0.07$, whereas the weighted mean for {\em Chandra}--SaWLens is $\langle b \rangle = -0.11 \pm 0.07$.  Some of the difference between these mass-bias measurements arises from a systematic offset between the WL and SaWLens mass profiles, which is $\approx 10 \pm 5$\% at 0.5~Mpc, averaged over all 16 clusters with both WL and SaWLens coverage (see Table~\ref{table:mass_bias}).  The remainder of the difference reflects the omission of two clusters with relatively large SaWLens--WL mass-ratio factors (Abell 383 and MS2137) from the {\em Chandra}--lensing comparison at 0.5~Mpc.

Combining the strong lensing with weak lensing (SaWLens) reduces the mass dispersion relative to {\em Chandra}, compared with weak lensing (WL) alone.  Dispersion in the mass ratio at 0.5~Mpc for the {\em Chandra}--SaWLens cluster set is $\lesssim 20$\% (see $\sigma_{\ln f}$ column of Table~\ref{table:mass_bias}), indicating an intrinsic scatter of $\lesssim 8$\% (see $\sigma_{\ln b}^{\rm intrinsic}$ column) after accounting for uncertainties in the mass-profile measurements.  
This quantity is not too different from what might be expected from the intrinsic scatter induced by projected structure along the line of sight.

As noted in Umetsu et al. (2014), the predictions of Rozo et al. (2014) for the mass enclosed within R$_{500}$ do not significantly differ from the weak-lensing masses in Umetsu et al. (2014):
$\langle M_{\rm{Rozo}}/M_{WL} \rangle = 1.13 \pm 0.10$. 
(To avoid aperture-induced errors, the weak lensing mass was computed inside the same radius.) 
Since Rozo et al. (2014) estimate a systematic uncertainty in their mass prediction (based on their X-ray luminosity estimates), there is no tension. The weak-lensing (and SaWLens) masses are what
Rozo et al.  would have expected for CLASH clusters based on their X-ray luminosities. Furthermore, the lensing-HSE X-ray mass ratio of $\sim 1$ is similar to what other groups have derived for Chandra HSE masses - weak lensing comparisons \citep{2013ApJ...767..116M,2013ApJ...765...24N,2014A&A...564A.129I}. For example, \citet{2013ApJ...767..116M} see this very same effect but apply a correction to the {\em Chandra} profiles to bring them into agreement with XMM. That work found a difference in the {\em Chandra}-XMM mass offset between non-cool core and cool-core clusters, in that their cool-core clusters had X-ray/weak-lensing mass ratios that were constant with radius while their non-cool core clusters had declining profiles like those in Figure~\ref{figure:massrat_xmm}. We do not see such a difference.  Also, when we add strong-lensing constraints to the lensing mass estimates, scatter in the X-ray/lensing mass relation decreases, indicating that the strong-lensing constraints improve the relation.

\section{XMM-Lensing Comparison} 
\label{section:xmm_lensing}

\begin{figure}[t]
  \begin{minipage}[b]{0.5\linewidth}
    \includegraphics[width=3.5in, trim = 0.6in 0.4in 0.6in 0.0in]{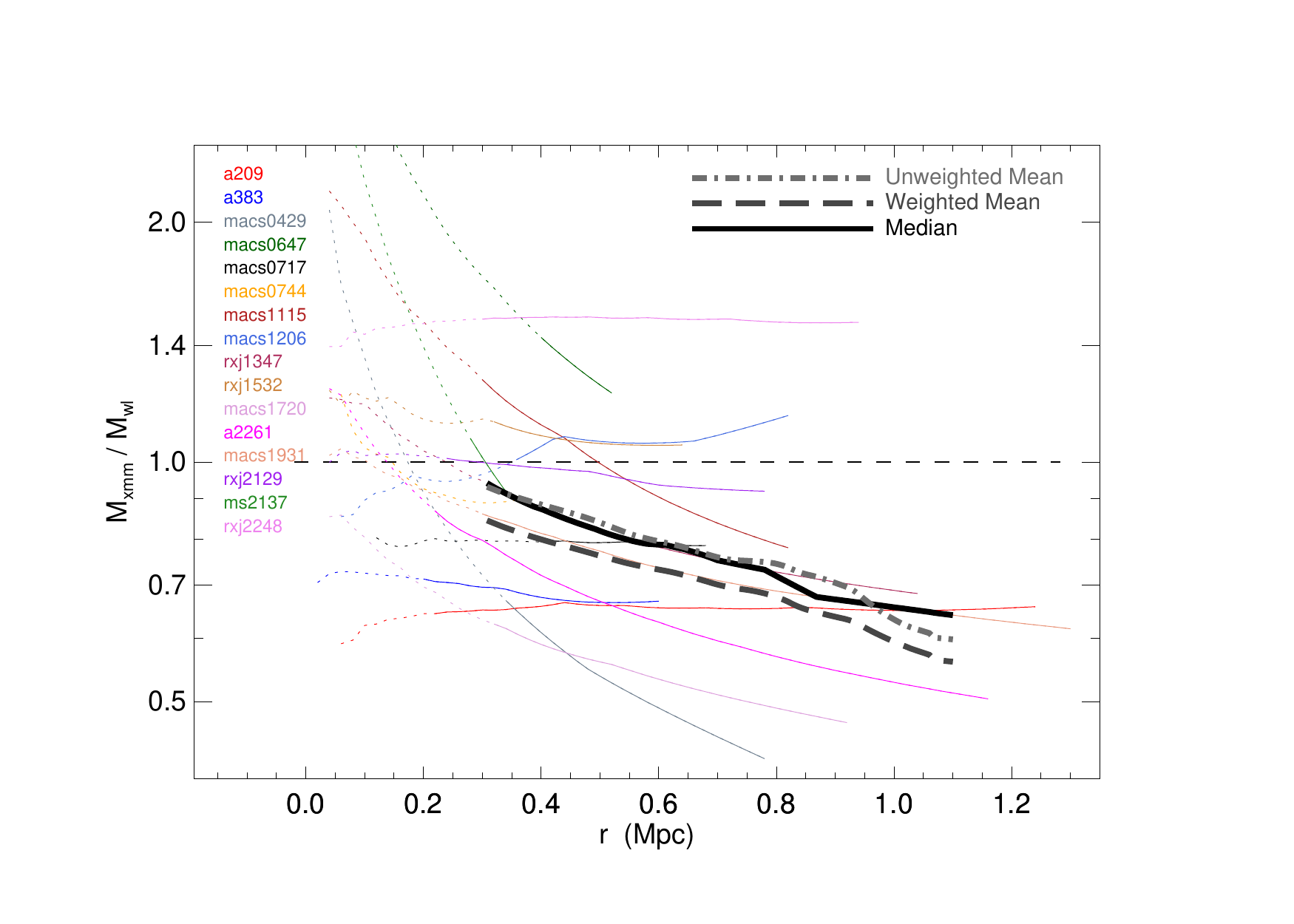}
  \end{minipage}
  \quad
  \begin{minipage}[b]{0.5\linewidth}
    \includegraphics[width=3.5in, trim = 0.6in 0.4in 0.6in 0.5in]{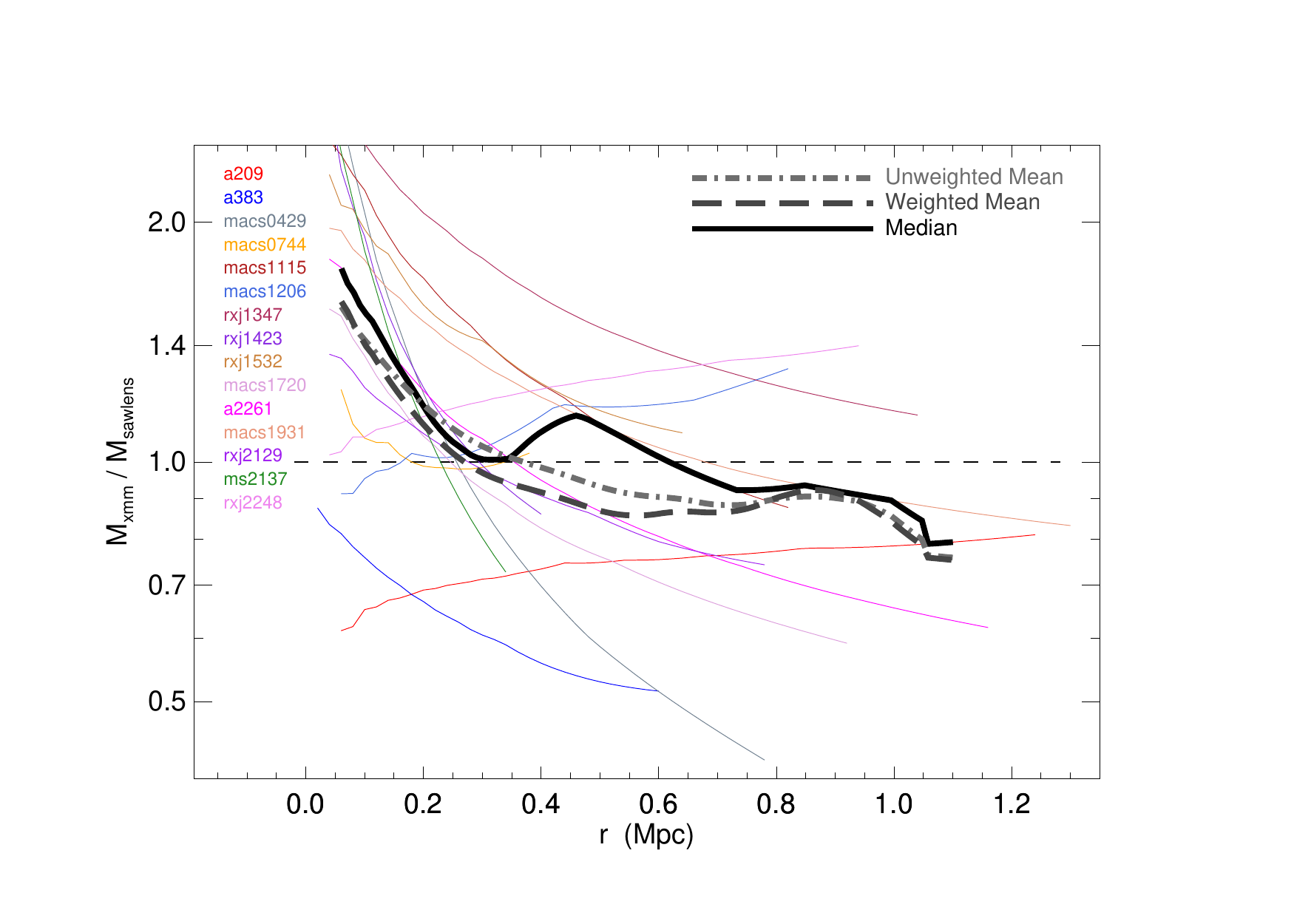} 
  \end{minipage}
  \caption{\footnotesize 
  Average ratios of JACO HSE mass profiles from XMM data to the CLASH weak-lensing (left panel) and 
  CLASH strong+weak lensing  (right panel) profiles.  
  Thick solid lines show the median ratios.  
  Long-dashed thick lines show weighted means.  
  Dot-dashed thick lines show unweighted means.  
  Short-dashed lines indicate the locus of equality.  
  Lists at left shows the clusters represented, whose best-fit profile ratios are given by the thin lines.
  Dotted extensions to those lines in the left panel show extrapolations inside of 1~arcminute, where the best-fitting
  NFW models to the WL data are not well constrained and are not used to compute means or medians.
  \label{figure:massrat_xmm}  
  }
\end{figure}

As one might anticipate from Figure~\ref{figure:T_MCMC}, our XMM mass-bias measurements are less well behaved than those for {\em Chandra}.  Figure~\ref{figure:massrat_xmm} shows how our XMM JACO mass profiles compare with those from the CLASH WL and SaWlens analyses.  There is a strong radial trend, with XMM masses tending to exceed lensing masses at small radii and to fall below them at large radii.  Also, the dispersions in mass bias for the XMM comparisons are greater than for the {\em Chandra} comparisons.  The weighted mean mass bias for XMM--WL at 0.5~Mpc is $\langle b \rangle = 0.24 \pm 0.05$ and for XMM--SaWlens is $\langle b \rangle = 0.18 \pm 0.08$.  It is smaller at 0.8~Mpc, because some of the more highly biased clusters have dropped out of the eight-cluster average at this radius, but this reversal of the overall trend appears to be a statistical fluke that does not represent the overall trend evident in the mass-profile ratios of individual clusters.  At 1.1~Mpc, the decline of the averages has resumed, but there are only three clusters remaining in the XMM sample at this radius, for which $\langle b \rangle = 0.44 \pm 0.05$ for XMM--WL and $\langle b \rangle = 0.25 \pm 0.07$ for XMM--SaWlens.  

Our results may seem to disagree with the general conclusions of \citet{2010ApJ...711.1033Z}, who find only a small mass discrepancy between weak lensing mass and XMM HSE masses. However, our X-ray XMM HSE masses agree quite well with theirs for the four CLASH clusters in common (see Table~\ref{table:Zhang}). We also note similar agreement for our XMM HSE mass estimates for these four clusters and the XMM estimates derived using independent analyses in \citet{2013A&A...550A.129P}.
The fact we derive very similar XMM HSE masses implies that our X-ray analyses are compatible with previous studies and  the weak lensing masses used in Zhang et al. and the {\em Planck} collaborationmust be somewhat lower than the lensing masses for CLASH clusters estimated by Umetsu et al. (2014) and Merten et al. (2014). 

In \S\ref{section:Planck_discrepancy} we discuss the implications of these findings for the discrepancy between the cosmological parameters inferred from the {\em Planck} analysis of primary CMB fluctuations and and the {\em Planck} S-Z cluster counts.

\begin{deluxetable}{lcccc}[b]
\tabletypesize{\scriptsize}
\tablecaption{LOCUSS (Zhang)-XMM comparison \label{table:Zhang}}
\tablewidth{0pt}
\tablehead{
\colhead{Name} & \colhead{Zhang M$_{2500}$ } & \colhead{Zhang $r_{S}$} & \colhead{CLASH M$_{2500}$} & \colhead{CLASH $r_{S}$} \\ 
 \colhead{}            & \colhead{ ($10^{14}$ M$_\odot$)} & \colhead{(Mpc)} & \colhead{($10^{14}$  M$_\odot$)} & \colhead{(Mpc)}    }\\
\startdata
A209  &$1.95\pm0.55$  &$ 0.208\pm0.014$ &  $1.73\pm0.11 $ &$0.22\pm0.005$ \\
A383  &$1.61\pm0.48$   &$ 0.127\pm0.007 $ &  $1.61\pm0.07 $ &$0.15\pm0.005$ \\
A2261 &$2.77\pm0.75$  &$ 0.266\pm0.026$ &  $2.78\pm0.09$ & $ 0.32\pm0.006$ \\
R2129 &$1.75\pm0.52$  &$ 0.166\pm0.009$ &  $1.96\pm0.64 $ & $0.24\pm0.004$ 
\enddata
\end{deluxetable}

\section{Gas Mass-Lensing Comparisons} 
\label{section:mgas_lensing}

At large radii, the total gas mass of a cluster is a potentially accurate proxy for its total mass.  Our CLASH-X profile comparisons support this notion (Figures ~\ref{figure:mgasmrat_chandra}-\ref{figure:mgasmrat_xmm}).  The fact that these mass-profile ratios generally flatten out near $\approx 0.125 = 1/8$ at $\sim 0.5$~Mpc (i.e., near $r_{2500}$ in these clusters) indicates that $8 M_{\rm gas}$ is a good mass proxy there, with only a small amount of bias (see also \citet{2003ApJ...590...15V,2004MNRAS.353..457A,2002MNRAS.334L..11A,2008MNRAS.387.1179M,2009A&A...501...61E}).  The flatness of the $M_{\rm gas}/M_{\rm SaWLens}$ profiles is particularly striking in this regard, especially for the {\em Chandra} comparison, since the formulae JACO uses to fit the gas density and dark-matter density profiles are completely different, and yet many of the ratio profiles are remaining within $\sim 10$\% of constancy from  $\lesssim 0.5$~Mpc to beyond $\gtrsim 1.0$~Mpc.  If this constancy in the ratio of gas mass to total mass continues to larger radii, it bodes well for the use of S-Z observations to measure total-mass profiles using gas mass as a proxy.

\begin{figure}[t]
  \begin{minipage}[b]{0.5\linewidth}
    \includegraphics[width=3.5in, trim = 0.6in 0.4in 0.6in 0.1in]{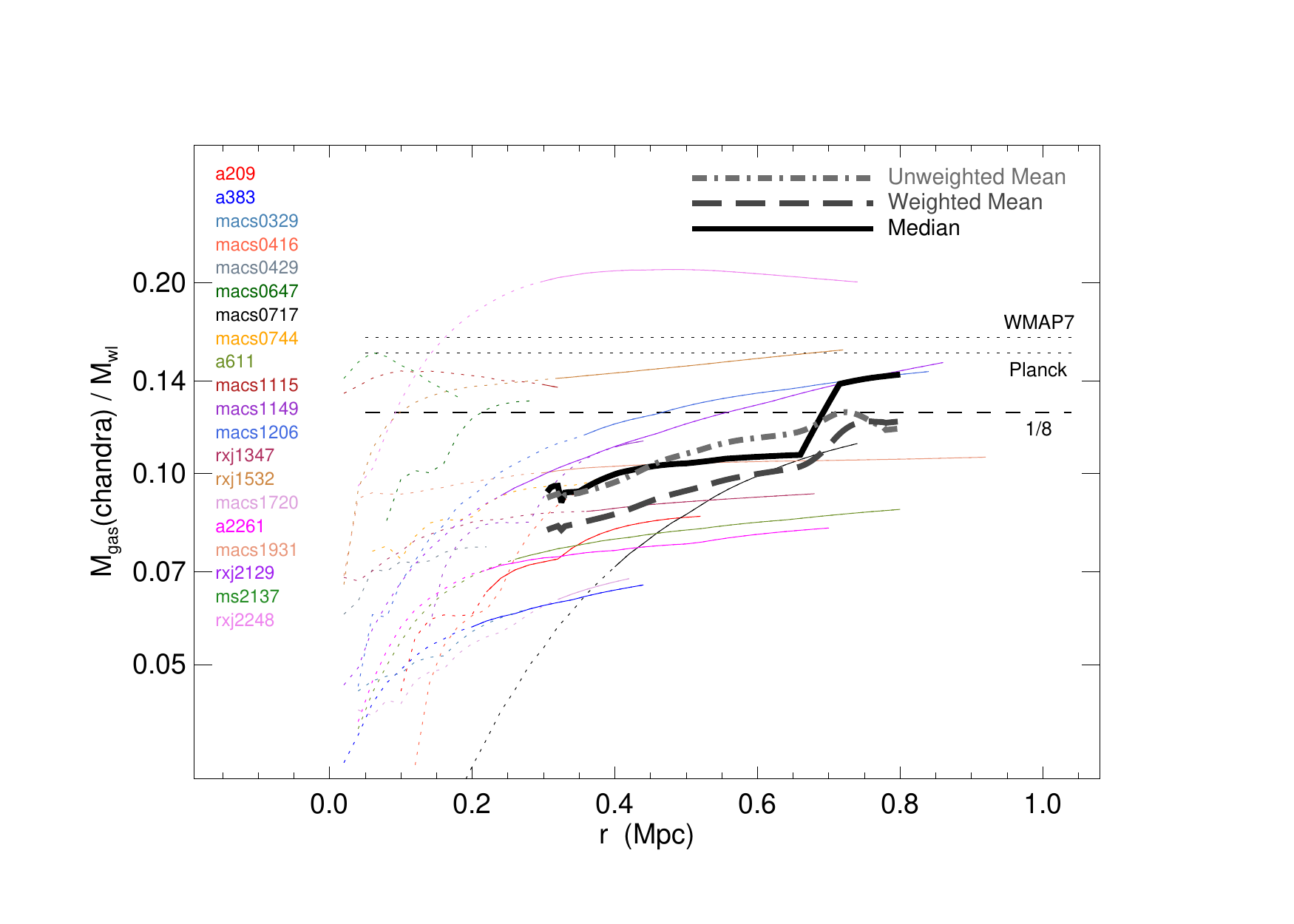}
  \end{minipage}
  \quad
  \begin{minipage}[b]{0.5\linewidth}
    \includegraphics[width=3.5in, trim = 0.6in 0.4in 0.6in 0.1in]{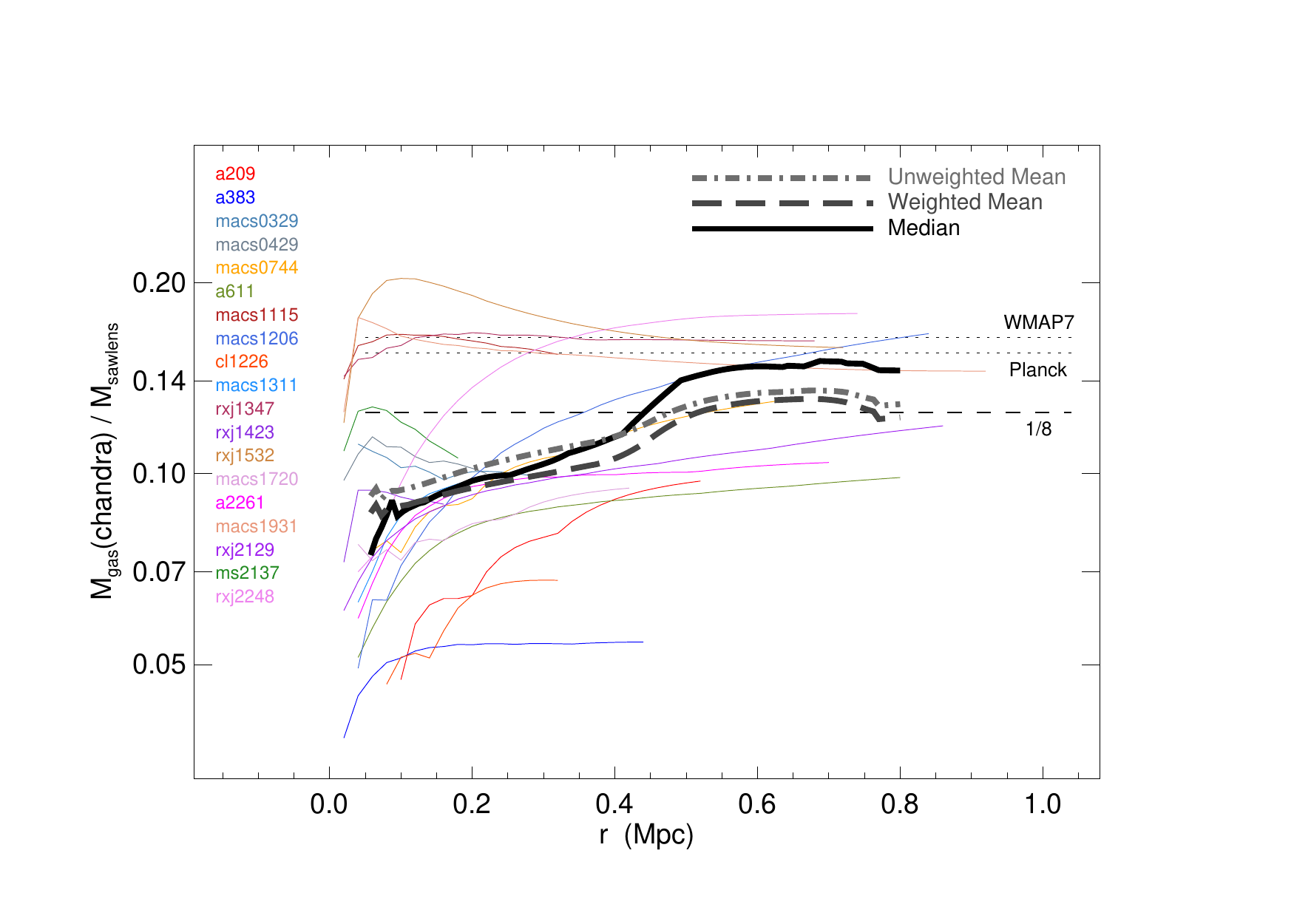} 
  \end{minipage}
  \caption{\footnotesize 
  Average ratios of {\em Chandra} gas-mass profiles to the CLASH weak-lensing (left panel) and 
  strong+weak lensing $M_{\rm gas}$ (right panel) profiles.  
  Thick solid lines show the median ratios.  
  Long-dashed thick lines show weighted means.  
  Dot-dashed thick lines show unweighted means.  
  Level short-dashed lines indicate a gas fraction of $1/8$.
  Level dotted lines show the cosmic baryon mass fractions found by {\em Planck} and WMAP.  
  Lists at left show the clusters represented, whose best-fit profile ratios are given by the thin lines.  
  Dotted extensions to those lines in the left panel show extrapolations inside of 1~arcminute, where the best-fitting
  NFW models to the WL data are not well constrained and are not used to compute means or medians.
  \label{figure:mgasmrat_chandra} 
  } 
\end{figure}

The bottom part of Table~\ref{table:mass_bias} lists average mass-ratio factors for comparisons of $M_{\rm gas}$ to lensing masses.  At 0.5~Mpc for {\em Chandra}, the weighted means imply $\langle f_{\rm gas} \rangle = 0.094 \pm 0.007$ relative to WL and $\langle f_{\rm gas} \rangle = 0.124 \pm 0.010$ relative to SaWLens.  At 0.5~Mpc for XMM, we find $\langle f_{\rm gas} \rangle = 0.093 \pm 0.006$ relative to WL and $\langle f_{\rm gas} \rangle = 0.117 \pm 0.011$ relative to SaWLens.  As radii rise to 1.1~Mpc, these gas fractions remain relatively constant.  Apparent differences between $\langle f_{\rm gas}$ values for {\em Chandra} and XMM do not arise from differences in the gas-mass measurements, because those are virtually identical (see Figure~\ref{figure:ne_mgas}).  Instead, they come from scatter in the lensing masses and the fact that the cluster-comparison sets for {\em Chandra} and XMM differ.

Notably, the intrinsic scatter between $M_{\rm gas}$ and lensing mass is smaller for SaWLens than for WL, in alignment with our finding for the comparisons of {\em Chandra} HSE masses with lensing masses.  Together, these findings confirm that the SaWLens analysis reduces the intrinsic scatter between true spherical mass and spherical mass inferred  from lensing, compared with the intrinsic scatter inferred from the weak-lensing data alone.  Furthermore, the intrinsic scatter for $M_{\rm gas}$ relative to SaWLens at 0.8~Mpc is only $\approx 12$\% for both XMM and {\em Chandra}, indicating that both $8M_{\rm gas}$ and SaWLens mass are low-scatter proxies for true spherical mass, with minimal bias.

At $f_{\rm gas}$ of 0.125, the hot gas in these massive 
clusters accounts for most but not all of the universal baryon budget of
0.155 \citep{2013arXiv1303.5076P}.  If the contents of clusters are, on average, representative of the rest of the universe, 
the remainder is likely to be largely made up of stars. Clusters could be preferentially baryon-poor compared to other regions in the universe, although depletions of
more than 10\% seem theoretically unlikely. However, a stellar mass fraction of 3\% is  
somewhat greater than is typically estimated for massive clusters, even when intracluster light is taken into account \citep{2013ApJ...778...14G}. 
But uncertainties in the BCG initial mass function and the systematic uncertainties in our mass estimates are still too great to conclude that these studies are missing much of the stellar mass.

\begin{figure}
  \begin{minipage}[b]{0.5\linewidth}
    \includegraphics[width=3.5in, trim = 0.6in 0.4in 0.6in 0.5in]{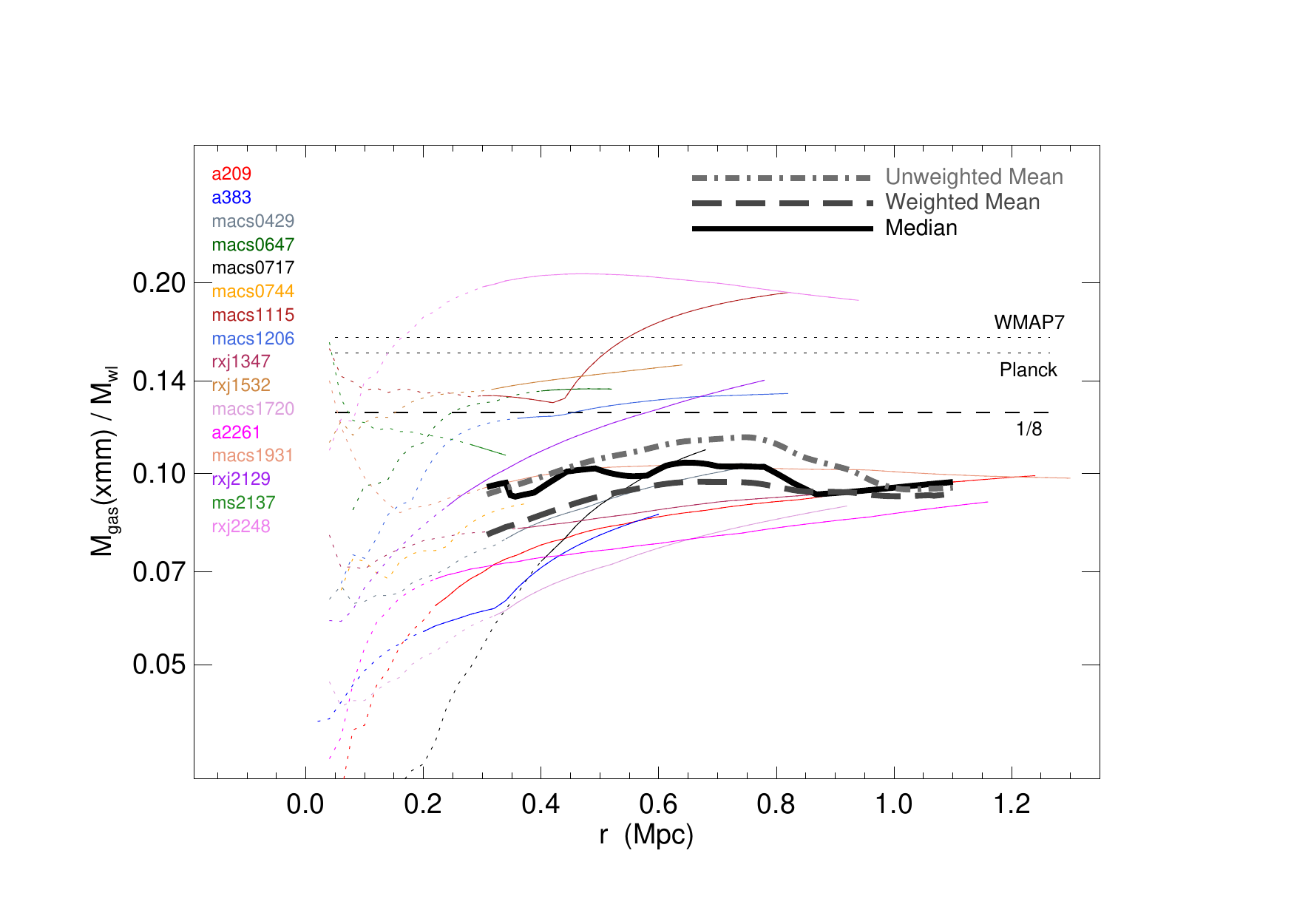}
  \end{minipage}
  \quad
  \begin{minipage}[b]{0.5\linewidth}
    \includegraphics[width=3.5in, trim = 0.6in 0.4in 0.6in 0.5in]{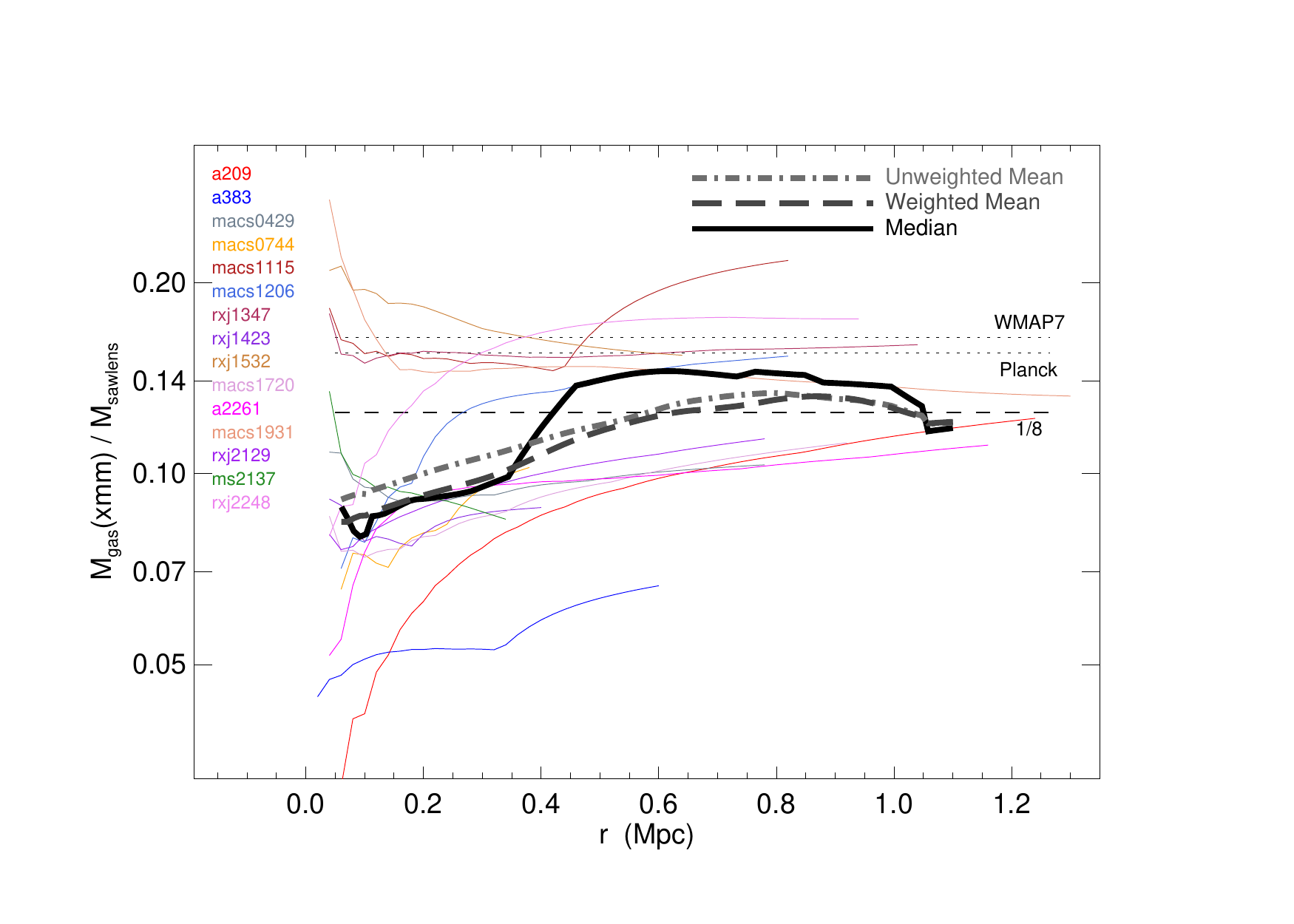} 
  \end{minipage}
  \caption{\footnotesize 
  Average ratios of {\em XMM} gas-mass profiles to the CLASH weak-lensing (left panel) and 
  strong+weak lensing  (right panel) profiles.  
  Thick solid lines show the median ratios.  
  Long-dashed thick lines show weighted means.  
  Dot-dashed thick lines show unweighted means.  
  Level short-dashed lines indicate a gas fraction of $1/8$.
  Level dotted lines show the cosmic baryon mass fractions found by {\em Planck} and WMAP.  
  Lists at left show the clusters represented, whose best-fit profile ratios are given by the thin lines.  
  Dotted extensions to those lines in the left panel show extrapolations inside of 1~arcminute, where the best-fitting
  NFW models to the WL data are not well constrained and are not used to compute means or medians.
  \vspace{1.0em}
  \label{figure:mgasmrat_xmm}
  }  
\end{figure}

\section{Implications for the {\em Planck} Cluster-Mass Discrepancy}
\label{section:Planck_discrepancy} 

In order to resolve the Planck Cluster-Mass discrepancy \citep{2013A&A...550A.131P}, that is, the tension 
between the number of clusters Planck finds via the Sunyaev-Zeldovich signal and 
the number of clusters predicted from 
the cosmological parameters inferred from the primary CMB power spectrum, a mass bias of 
$\langle b_{\rm XMM} \rangle \sim 0.4$ \citep{2013A&A...550A.131P} is needed. 
Our comparison between XMM and lensing masses for the CLASH sample does indeed find significant mass bias 
in XMM hydrostatic mass measurements. However, assigning a single number to that bias is difficult, because it
depends both on the radius within which mass is measured and the lensing data (i.e. 
weak lensing or SaWLens) used to determine the magnitude of that bias. 
Additionally, the XMM-WL bias of $\langle b_{\rm XMM} \rangle \sim 0.44$ at 1.1 $h_{70}^{-1}$ Mpc is sufficient to
account for the entire discrepancy, but is based on averaging over only three CLASH clusters.  
 
Alternatively, one can extrapolate the best-fitting JACO X-ray mass profiles to radii larger than the range of the hydrostatic
model.  All the CLASH clusters with lensing data can then be included in the averages, but at the expense of statistical significance
and perhaps also additional systematic biases.  With these caveats, we present in Table~5 the mass bias factors obtained
from extrapolations out to $r_{500}$, using the $r_{500}$ determined from the same data set.  That procedure 
further amplifies any mass bias present at a fixed radius, because of the aperture-induced covariance effect, and
results in $\langle b_{\rm XMM-WL} \rangle = 0.44 \pm 0.06$ and $\langle b_{\rm XMM-SaWlens} \rangle = 0.47 \pm 0.08$.
The $\chi_\nu^2$ values for these averages are unacceptably large, indicating that we have stretched most of the CLASH
X-ray data beyond the limits of reliability, meaning that the formal uncertainties on these mass-bias values are too small.  
However, the overall trend does appear to be real.

Our results therefore are in alignment with the finding of the Weighing the Giants collaboration \citep{2014arXiv1402.2670V}, 
that the default mass calibration adopted by the Planck team ($\langle b \rangle = 0.2$) underestimates the true masses at large radii. 
We approached this question somewhat differently from the WtG team, in that we are deriving HSE masses directly from the XMM data 
for our sample of  CLASH clusters, and because our mass profiles benefit from additional information:  
Merten et al. (2014) utilize strong lensing constraints on the weak shear profiles in the case of SaWLens and 
Umetsu et al. (2014) include magnification constraints in the case of CLASH-WL. These new lensing masses are consistent with the lensing masses
derived by WtG based on shear alone; our analysis of the X-ray observations for the same clusters show that the XMM HSE masses,
derived directly (and not from scaling relations) are also consistent with $\langle b \rangle $ considerably larger than 0.2, for
either WL or SaWLens masses as surrogates for the gravitating masses.

\section{Conclusions}

These are our primary findings:

\begin{enumerate}

\item {\em Chandra} and XMM measurements of electron density and enclosed gas mass as functions of radius are highly consistent with one another, indicating that any differences in HSE masses inferred from X-ray observations arise from differences in gas-temperature measurements (\S\ref{section:ne_profiles}).

\item Gas temperatures measured in clusters by XMM and {\em Chandra} are consistent with one another at $\sim 100$~kpc radii but XMM temperatures systematically decline relative to {\em Chandra} temperatures as the radius of the temperature measurement increases (\S\ref{section:T_profiles}).  Plausible contributions to this apparent temperature difference are large-angle scattering of soft X-ray photons in excess of that amount expected from the standard XMM PSF correction, a radial variation in the quality of the soft energy calibration of both telescopes, and uncertain vignetting corrections. While we cannot state with finality that the Chandra absolute calibration is better than XMM's, the {\em Chandra}-derived cluster HSE mass profiles are significantly more similar in shape and normalization to the CLASH strong and weak lensing profiles presented in Umetsu et al. (2014) and Merten et al. (2014). We plan and encourage future work in cross-comparison of XMM and Chandra cluster analyses to unlock the full potential of the investment of both observatories.

\item We present the CLASH-X mass-profile comparisons in the form of circular-velocity profiles, because sharing and comparing results in that form has several advantages:  Mass profiles provided in terms of $v_{\rm circ}(\theta_r)$ are independent of cosmological assumptions.  Plots of $v_{\rm circ}(r)$ span much less dynamic range than $M_r$ plots, making systematic differences among profiles more apparent.   The scale radius $r_{\rm s}$ and maximum circular velocity $v_{\rm max}$ of a halo do not change if the halo does not change, whereas its mass, radius, and concentration continually increase if those quantities are defined with respect to a spherical-overdensity threshold $\Delta$.  A halo's value of $v_{\rm max}$ is therefore a more general indicator of its properties than $M_\Delta$, because it is independent of redshift.  Furthermore, accurate estimates of $v_{\rm max}$ can be obtained from information at many different radii, because $v_{\rm circ}(r)$ curves for NFW profiles are nearly level at the radii of greatest interest, remaining within 2\% of $v_{\rm max}$ over the interval $1.4 \lesssim r / r_{\rm s} \lesssim 3.5$.

\item Ratios of {\em Chandra} HSE mass profiles to CLASH strong and weak lensing profiles show no obvious radial dependence in the 0.3--0.8 Mpc range.  However, the mean mass biases inferred from the WL and SaWLens data are different, with a weighted-mean value at 0.5 Mpc of $\langle b \rangle = 0.12$ for the WL comparison and $\langle b \rangle = -0.11$ for the SaWLens comparison.

\item Ratios of XMM HSE mass profiles to CLASH lensing profiles show a pronounced radial dependence in the 0.3--1.0 Mpc range, with a weighted-mean mass bias of value rising to $\langle b \rangle = 0.3$ at 1~Mpc for the WL comparison and $\langle b  \rangle = 0.2$ for the SaWLens comparison.

\item Enclosed gas mass profiles from both {\em Chandra} and XMM rise to $\sim 0.125$ times the total-mass profiles inferred from lensing at $\approx 0.5$~Mpc and remain constant outside of that radius, indicating that $8 M_{\rm gas}$ profiles may be a useful proxy for total-mass profiles at $\gtrsim 0.5$~Mpc in massive galaxy clusters.

\end{enumerate}

\acknowledgments

MD and AH acknowledge the partial support of STScI/NASA award HST-GO-12065.07-A and NASA award NNX13AI41G. 
MD and GMV benefitted from discussions of this work with Jim Bartlett.
The Dark Cosmology Centre is funded by the DNRF. AM was partially supported through NASA ADAP award NNX12AE45G. 
SE acknowledges the financial contribution from contracts ASI-INAF I/023/05/0 and I/088/06/0.
The work of LAM and JM was carried out at Jet Propulsion Laboratory,
California Institute of Technology, under a contract with NASA. 
MM acknowledges financial contribution from the agreement ASI/INAF I/023/12/0 and from the project INFN PD51.
Support for AZ is provided by NASA through a Hubble Fellowship grant HST-HF-51334.01-A awarded by STScI.

{\it Facilities:} \facility{Chandra X-ray Observatory} \facility{Newton X-ray Multi Mirror Mission} \facility{Hubble Space Telescope}

\vspace{1em}
\bibliography{CLASH}

\end{document}